\documentclass[twocolumn]{aastex63}

\usepackage{xspace}
\usepackage{amsmath}
\usepackage{booktabs}

\usepackage{xstring} 

\newcommand{\constantperiodmodel}[1]{%
  \IfEqCase{#1}{%
  {per}{$1.091419649 \pm 0.000000025$~days}
  {per_val}{1.091419649}
  {per_unc}{25}
  {t0}{$2456305.455521^{+0.000026}_{-0.000025}$}
  {t0_val}{2456305.455521}
  {t0_unc}{26}
  {ndof}{156}
  {chisq}{380.745}
  {chisq_rounded}{380.7}
  {bic}{390.871}
  {bic_rounded}{390.9}
  }[\PackageError{tree}{Undefined option to tree: #1}{}]
}

\newcommand{\decaymodel}[1]{%
  \IfEqCase{#1}{%
  {per}{$1.091420107 \pm 0.000000042$~days}
  {per_val}{1.091420107}
  {per_unc}{42}
  {t0}{$2456305.455809 \pm 0.000032$}
  {t0_val}{2456305.455809}
  {t0_unc}{32}
  {pdot}{$-9.20^{+0.62}_{-0.63} \times 10^{-10}$~days/day}
  {pdot_yr}{$-29 \pm 2$~ms/yr}
  {pdot_val}{$-9.20$}
  {pdot_unc}{62}
  {dpde}{$-10.0^{+0.68}_{-0.69} \times 10^{-10}$~days/epoch}
  {dpde_val}{$-10.04$}
  {dpde_unc}{$69$}
  {ndof}{155}
  {chisq}{167.566}
  {chisq_rounded}{167.6}
  {bic}{182.754}
  {bic_rounded}{182.8}
  }[\PackageError{tree}{Undefined option to tree: #1}{}]
}%

\newcommand{\precessmodel}[1]{%
  \IfEqCase{#1}{%
  {per}{$1.091419633 \pm 0.000000081$~days}
  {per_val}{1.091419633}
  {per_unc}{81}
  {t0}{$2456305.45488 \pm 0.00012$}
  {t0_val}{2456305.45488}
  {t0_unc}{12}
  {e}{$0.00310^{+0.00035}_{-0.00034}$}
  {e_val}{0.00310}
  {e_unc}{$35$}
  {w}{$2.62 \pm 0.10$~rad}
  {w_val}{2.62}
  {w_unc}{10}
  {wdot}{$0.000984 ^{+0.000070}_{-0.00061}$~rad/day}
  {wdot_val}{0.000984}
  {wdot_unc}{$+70,-61$}
  {ndof}{153}
  {chisq}{179.700}
  {chisq_rounded}{179.7}
  {bic}{205.013}
  {bic_rounded}{205.0}
  }[\PackageError{tree}{Undefined option to tree: #1}{}]
}%

\newcommand{\timingcomparison}[1]{%
  \IfEqCase{#1}{%
  {dbic}{22.3}
  {dchisq}{12.1}
  }[\PackageError{tree}{Undefined option to tree: #1}{}]
}%

\newcommand{\rvmodel}[1]{%
  \IfEqCase{#1}{%
  {per}{$1.0914177 \pm 0.0000023$}
  {tc}{$2454508.9801 \pm 0.0033$}
  {K}{$218 \pm 2$}
  {dvdt}{$0.0003 \pm 0.0024$}
  {gamma_harps}{$19087.8 \pm 3.7$}
  {jit_harps}{$11.8 \pm 3.3$}
  {gamma_hires}{$12.8\pm 2.3$}
  {jit_hires}{$9.9 \pm 1.3$}
  {gamma_sophie}{$19083.7 \pm 5.3$}
  {jit_sophie}{$16.1 \pm 2.8$}
  }[\PackageError{tree}{Undefined option to tree: #1}{}]
}%

\newcommand{\rveccmodel}[1]{%
  \IfEqCase{#1}{%
  {per}{$1.0914207 \pm 0.0000023$}
  {tc}{$2454508.9765 \pm 0.0039$}
  {K}{$219 \pm 2$}
  {secosw}{$-0.0004 \pm 0.029$}
  {sesinw}{$-0.175^{+0.028}_{-0.024}$}
  {e}{$0.0317 \pm 0.0087$}
  {w}{$-89.9^{+9.7}_{-9.2}$~deg}
  {dvdt}{$-0.0005 \pm 0.0023$}
  {gamma_harps}{$19089.0 \pm 3.4$}
  {jit_harps}{$9.7^{+3.4}_{-2.6}$}
  {gamma_hires}{$13.3\pm 2.2$}
  {jit_hires}{$8.9^{+1.4}_{-1.1}$}
  {gamma_sophie}{$19080.7 \pm 5.4$}
  {jit_sophie}{$15.3^{+3.2}_{-2.5}$}
  }[\PackageError{tree}{Undefined option to tree: #1}{}]
}%

\newcommand{\val}[2]{%
  \edef\x{#2_val}
  \unskip
  \IfEqCase{#1}{%
  	{constant-period}{\constantperiodmodel{\x}}%
    {decay}{\decaymodel{\x}}%
    {precess}{\precessmodel{\x}}%
  }[\PackageError{tree}{Undefined option to tree: #1}{}]%
}%

\newcommand{\unc}[2]{%
  \edef\x{#2_unc}
  \unskip
  \IfEqCase{#1}{%
    {constant-period}{\constantperiodmodel{\x}}%
    {decay}{\decaymodel{\x}}%
    {precess}{\precessmodel{\x}}%
  }[\PackageError{tree}{Undefined option to tree: #1}{}]%
}%

\newcommand{\rounded}[2]{%
  \edef\x{#2_rounded}
  \unskip
  \IfEqCase{#1}{%
    {constant-period}{\constantperiodmodel{\x}}%
    {decay}{\decaymodel{\x}}%
    {precess}{\precessmodel{\x}}%
  }[\PackageError{tree}{Undefined option to tree: #1}{}]%
}%

\newcommand{\Spitzer}{\textit{Spitzer}\xspace}
\newcommand{\TESS}{\textit{TESS}\xspace}

\newcommand{\bjdtdb}{\ensuremath{\mathrm{BJD}_\mathrm{TDB}}\xspace}
\newcommand{\Rp}{\ensuremath{R_\mathrm{P}}\xspace}
\newcommand{\Mp}{\ensuremath{M_\mathrm{P}}\xspace} 
\newcommand{\Rstar}{\ensuremath{R_{\star}}\xspace} 

\newcommand{\Rjup}{\ensuremath{R_\mathrm{J}}\xspace} 
\newcommand{\Mjup}{\ensuremath{M_\mathrm{J}}\xspace}
 
\newcommand{\Msun}{\ensuremath{M_\odot}\xspace}

\newcommand{\Qs}{\ensuremath{Q_\star'}\xspace}
\newcommand{\Teff}{\ensuremath{T_{\mathrm{eff}}}\xspace}
\newcommand{\feh}{\ensuremath{\mathrm{[Fe/H]}}\xspace}
\newcommand{\logg}{\ensuremath{\log g}\xspace} 

\newcommand{\chisq}{\ensuremath{\chi^2}\xspace}
\newcommand{\dchisq}{\ensuremath{\Delta\chi^2}\xspace}
\newcommand{\chisqmin}{\ensuremath{\chi^2_\mathrm{min}}\xspace}
\newcommand{\dbic}{\ensuremath{\Delta\mathrm{BIC}}\xspace}

\newcommand{\um}{\ensuremath{\mu \mathrm{m}}\xspace}
\newcommand{\ms}{\ensuremath{\mathrm{m}\,\mathrm{s}^{-1}}\xspace}

\newcommand{\msday}{\ensuremath{\mathrm{m}\,\mathrm{s}^{-1}\,\mathrm{day}^{-1}}\xspace}
\newcommand{\msyr}{\ensuremath{\mathrm{m}\,\mathrm{s}^{-1}\,\mathrm{yr}^{-1}}\xspace}

\newcommand{\lonperi}{\ensuremath{\omega_{\star}}\xspace}

\newcommand{\secosw}{\ensuremath{\sqrt{e} \cos \lonperi}\xspace}
\newcommand{\sesinw}{\ensuremath{\sqrt{e} \sin \lonperi}\xspace}

\newcommand{\sigjit}{\ensuremath{\sigma_\mathrm{jit}}\xspace}

\begin{document}
\title{The Orbit of WASP-12b is Decaying}

\correspondingauthor{Samuel W.\ Yee}
\email{swyee@princeton.edu}

\author[0000-0001-7961-3907]{Samuel W.\ Yee}
\author[0000-0002-4265-047X]{Joshua N.\ Winn}
\affiliation{Department of Astrophysical Sciences, Princeton University, 4 Ivy Lane, Princeton, NJ 08540, USA}

\author{Heather A.\ Knutson}
\affiliation{Division of Geological and Planetary Sciences, California Institute of Technology, 1200 California Blvd, Pasadena, CA 91125, USA}
\author{Kishore C.\ Patra}
\affiliation{Department of Astronomy, University of California, Berkeley, CA 94720, USA}
\author[0000-0003-2527-1475]{Shreyas Vissapragada}
\author[0000-0002-0659-1783]{Michael M.\ Zhang}
\affiliation{Division of Geological and Planetary Sciences, California Institute of Technology, 1200 California Blvd, Pasadena, CA 91125, USA}
\author[0000-0002-1139-4880]{Matthew J.\ Holman}
\affiliation{Harvard-Smithsonian Center for Astrophysics, 60 Garden Street, Cambridge, MA 02138, USA}
\author[0000-0002-1836-3120]{Avi Shporer}
\affiliation{Department of Physics and Kavli Institute for Astrophysics and Space Research, Massachusetts Institute of Technology, Cambridge, MA 02139, USA}
\author[0000-0001-6160-5888]{Jason T.\ Wright}
\affiliation{Department of Astronomy \& Astrophysics, 525 Davey Laboratory, The Pennsylvania State University, University Park, PA, 16802, USA}
\affiliation{Center for Exoplanets and Habitable Worlds, 525 Davey Laboratory, The Pennsylvania State University, University Park, PA, 16802, USA}

\begin{abstract}
WASP-12b is a transiting hot Jupiter on a 1.09-day orbit around a late-F star.
Since the planet's discovery in 2008, the time interval between transits has been decreasing by $29\pm 2$ msec~year$^{-1}$.
This is a possible sign of orbital decay, although the  previously available data left open the possibility that the planet's orbit is slightly eccentric and is undergoing apsidal precession.
Here, we present new transit and occultation observations that provide more decisive evidence for orbital decay, which is favored over apsidal precession by a \dbic of \timingcomparison{dbic} or Bayes factor of 70,000.
We also present new radial-velocity data that rule out the R{\o}mer effect as the cause of the period change.
This makes WASP-12 the first planetary system for which we can be confident that the orbit is decaying.
The decay timescale for the orbit is $P/\dot{P} = 3.25\pm 0.23$~Myr.
Interpreting the decay as the result of tidal dissipation,
the modified stellar tidal quality factor is $Q'_\star = 1.8 \times10^{5}$.
\end{abstract}

\section{Introduction} \label{sec:intro}

There are several reasons why the orbital period of a hot Jupiter might change, or appear to change.
Interactions with other planets cause transit-timing variations, although it is now well established that hot Jupiters tend to lack planetary companions close
enough or massive enough to produce detectable variations (see, e.g., \citealt{Steffen2012,Huang2016}).
On secular timescales, a planetary or stellar companion can induce orbital precession
\citep{Miralda-Escude2002} or Kozai-Lidov cycles \citep{Holman1997,Innanen1997,Mazeh1997}.
Even in the absence of external perturbers, an eccentric orbit will precess due to general relativity and the quadrupole fields from rotational and tidal bulges \citep{Jordan2008,Pal2008}.
There are also the long-term effects of tidal dissipation, which for hot Jupiters are expected to lead to orbital circularization, coplanarization, and decay \citep{Counselman1973,Hut1980,Rasio1996,Levrard2009}.
Mass loss might cause the orbit to expand or contract, depending on the specific angular momentum of the escaping material and where it is ultimately deposited (see, e.g., \citealt{Valsecchi2015,Jackson2016}).
Finally, any long-term acceleration of the host star will cause an illusory change in period due to the associated changes in the light-travel time, known as the R{\o}mer effect.
Such an acceleration would likely be due to a wide-orbiting companion.

Of all these possibilities, the most interesting are probably tidal orbital decay, mass loss, and apsidal precession, because the measured rate of change would give us insight into a poorly understood phenomenon.
The rate of tidal orbital decay depends on the unknown mechanisms by which the stellar tidal oscillations are dissipated as heat \citep{Rasio1996,Sasselov2003}.
Mass loss could be due to an escaping wind, or Roche lobe overflow, either of which could
be precipitated by tidal orbital decay \citep{Valsecchi2015}.
The rate of apsidal precession is expected to be dominated by the contribution from the planet's tidal deformability, and therefore, the measured rate would give us a glimpse into the planet's interior structure \citep{Ragozzine2009}.

Currently, the most promising system for observing these effects is WASP-12 \citep{Hebb2009,Haswell2018}.
The host star is a late-F star ($\Teff \approx 6300\,{\rm K}$; \citealt{Hebb2009}). The planet is a hot Jupiter with orbital period 1.09 days, mass $1.47$~\Mjup, and radius $1.90$~\Rjup \citep{Collins2017}.
This radius is unusually large even by the standards of hot Jupiters, and ultraviolet transit observations imply an even larger cloud of diffuse gas, indicating that the planet has an escaping exosphere
\citep{Fossati2010,Haswell2012,Nichols2015}.
Furthermore, there is evidence for variations in the time interval between transits.
\cite{Maciejewski2011} reported the detection of short-timescale variations, although subsequent analysis by \cite{Maciejewski2013} showed that the statistical significance was weaker than originally reported.
\cite{Maciejewski2013} also presented a larger database of transit times and found evidence that the interval between transits is varying sinusoidally with a 500-day period.
They hypothesized that the anomalies were due to a second planet in the system with a mass of 0.1~\Mjup and a period of 3.5~days.

After accumulating more data, \cite{Maciejewski2016} did not confirm the sinusoidal variability, but instead found a quadratic trend consistent with a uniformly decreasing orbital period.
\cite{Patra2017} presented new data and confirmed that the observed interval between transits has been decreasing, at a rate of $29\pm 3$~msec~yr$^{-1}$.
\cite{Patra2017} also showed that the available radial-velocity data were incompatible with a line-of-sight acceleration large enough for the R{\o}mer effect to be the sole explanation for the apparent decrease in orbital period. Additional transit times have since been reported by \cite{Maciejewski2018} and \cite{Baluev2019}, in both cases supporting the finding of a long-term decrease in the transit period.

\cite{Bailey2019} considered and discarded many explanations for the period decrease besides tidal
orbital decay, such as the Applegate effect or gravitational perturbations from another planet.
However, the possibility remained that the orbit is eccentric and apsidally precessing, and that the apparently quadratic trend in the transit timing deviations is really a portion of a long-period sinusoidal pattern.
The radial-velocity data rule out eccentricities larger than about 0.03, 
but even an eccentricity on the order of $10^{-3}$ would be sufficient to fit the data under this hypothesis.

One way to tell the difference between orbital decay and apsidal precession is to measure the times of occultations (secondary eclipses).
In the case of orbital decay, the time interval between occultations would be shrinking at the same rate as the time interval between transits.
In contrast, for a precessing orbit, the transit and occultation timing deviations would have opposite signs.
\cite{Patra2017} analyzed all of the available occultation times and found that both models gave a reasonable fit to the data.
They found a preference for orbital decay over apsidal precession, but because the statistical significance was modest ($\Delta\chi^2 = 5.5$), they stopped short of claiming conclusive evidence for orbital decay.
By extrapolating both models into the future, \cite{Patra2017} showed that observations of occultations
over the next few years would allow for a more definitive conclusion.

Two years have now elapsed.  In this paper, we 
report new observations of transits (Section \ref{sec:transits}) and
occultations (Section \ref{sec:occultations}),
as well as additional radial-velocity data (Section \ref{sec:rvs}).
We present an analysis of all the available data,
finding that orbital decay is favored over apsidal precession with greater
confidence than before (Section \ref{sec:analysis}).
Finally, we discuss the possible implications of the observed decay rate
for our understanding of hot Jupiters and stellar interiors (Section \ref{sec:discussion}).

\section{New Transit Observations} \label{sec:transits}

We observed ten transits of WASP-12b with the 1.2m telescope at the Fred Lawrence Whipple Observatory (FLWO) on Mt.\ Hopkins, Arizona, between November 2017 and January 2019.
The observations were made with {\it Keplercam} and a Sloan \textit{r'}-band filter, with an exposure time of 15 seconds,
yielding a typical signal-to-noise ratio of 200 per frame.
We reduced the data with standard procedures, as described by \cite{Patra2017}.
We performed circular-aperture photometry of WASP-12 and 7--9 comparison stars.
The aperture radius was typically 7-8 pixels, chosen to minimize the scatter in the out-of-transit flux of WASP-12
relative to the comparison stars.
We then produced light curves by dividing the flux of WASP-12
by the sum of the comparision star fluxes, and then normalizing to set the median flux equal to unity outside
of the transits.
The estimated uncertainty in each data point was taken to be the standard deviation of the flux time series outside of
transits.
\added{The photometric time series is provided in Table \ref{tab:phot_data}.}

\begin{figure}
\plotone{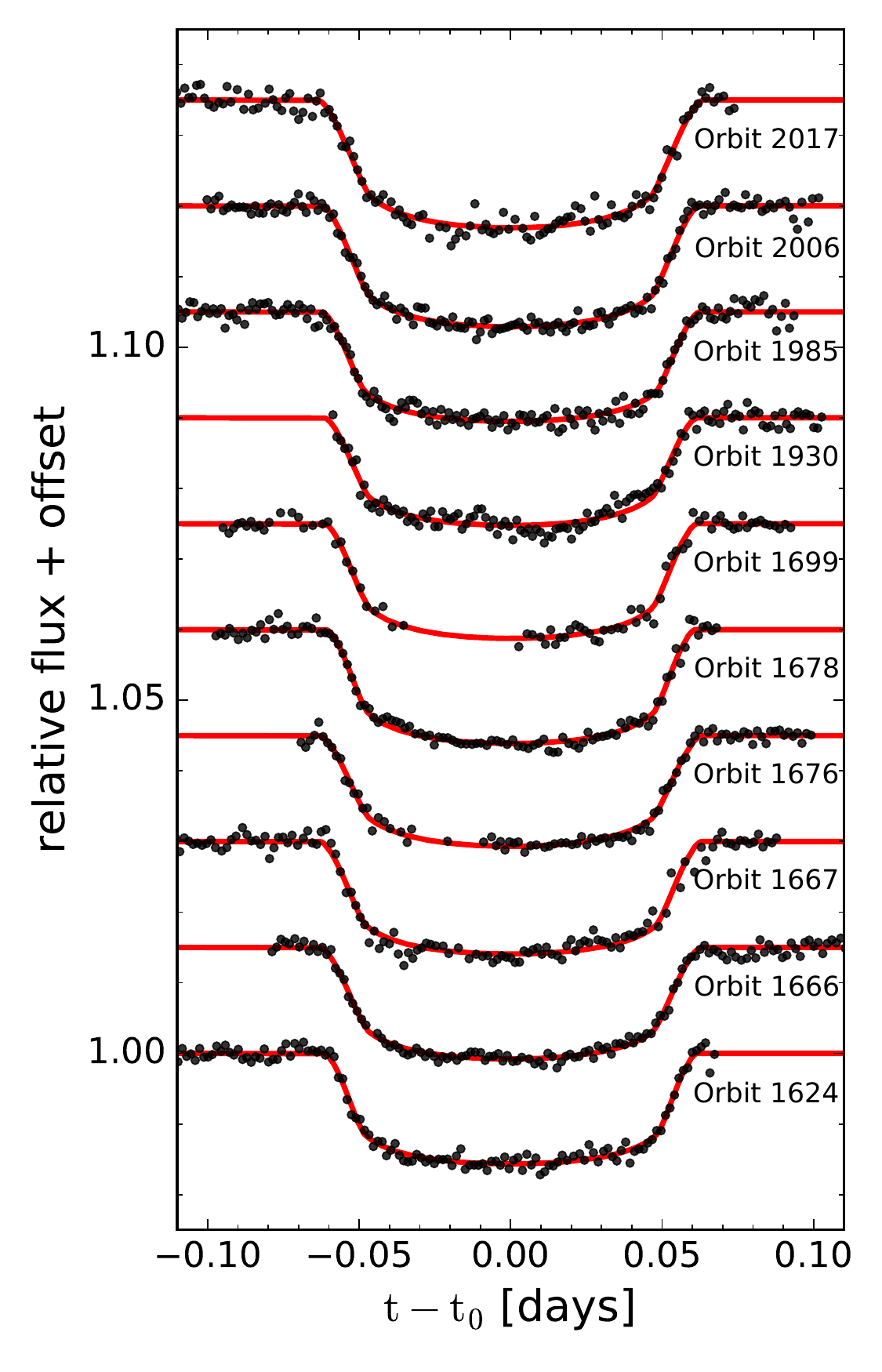}
\caption{Transit light curves of WASP-12b, based on $r'$-band observations with the FLWO 1.2m telescope.
The red curves are based on the best-fitting model.
The number on the right side of each light curve is the orbit number relative to an fixed reference orbit.
\label{fig:transit_lightcurves}}
\end{figure}

To measure transit times, we fitted a standard transit model \citep{Mandel02}.
\replaced{We assumed the limb-darkening function to be quadratic, with
coefficients $u_1=0.32$ and $u_2 = 0.32$, based on the tables provided by \cite{Claret2011}.}
{We assumed a quadratic limb-darkening law with coefficients $u_1 = 0.32, u_2 = 0.32$, as tabulated by \cite{Claret2011} for a star having the spectroscopic properties $\Teff = 6290~\mathrm{K}, \feh = 0.3, \logg = 4.3$ \citep{Hebb2009}\footnote{Here, we used the online tool of \cite{Eastman2013} at \url{http://astroutils.astronomy.ohio-state.edu/exofast/limbdark.shtml} to interpolate the \cite{Claret2011} tables.}.}
We obtained the best fit to each light curve by minimizing the usual \chisq statistic.
We then used the \texttt{emcee} code \citep{Foreman-Mackey13,Goodman10} to perform an affine-invariant Markov Chain Monte Carlo (MCMC) sampling to determine the uncertainties in the model parameters, including the transit time (the midpoint of the transit, or the time of minimum light).
\added{We discarded the first $\sim 30\%$ of the MCMC chains as burn-in, and ensured convergence by comparing chains from multiple MCMC runs with the Gelman-Rubin statistic \citep{GelmanRubin}.
}
Figure \ref{fig:transit_lightcurves} shows the new light curves and the best-fitting model curves.
Table \ref{tab:timing_data} gives the transit times and uncertainties.
The typical uncertainty is 30~seconds, comparable to the precision obtained by \cite{Patra2017}.


\begin{deluxetable}{lccc}
\tablecolumns{4}
\tablewidth{\columnwidth}
\tablecaption{Photometric timeseries \label{tab:phot_data}}
\tablehead{
	\colhead{\bjdtdb} & \colhead{Normalized Flux} & \colhead{$\sigma(\mathrm{Flux})$} & \colhead{Code\tablenotemark{a}}
}
\startdata
2458123.68083 & 0.9980 & 0.0014 & F1666 \\
2458123.68118 & 0.9985 & 0.0014 & F1666 \\
2458123.68153 & 0.9991 & 0.0014 & F1666 \\
2458123.68192 & 1.0000 & 0.0014 & F1666 \\
2458123.68233 & 0.9992 & 0.0014 & F1666 \\
2458123.68270 & 1.0000 & 0.0014 & F1666 \\
2458123.68305 & 0.9997 & 0.0014 & F1666 \\
2458123.68340 & 0.9993 & 0.0014 & F1666 \\

\enddata
\tablenotetext{a}{Code denotes the source and orbit number for each data point. The first character represents the source telescope -- F for FLWO transit observations, S for \Spitzer occultation observations, W for WIRC occultation observations.}
\tablecomments{Table \ref{tab:phot_data} is published in its entirety in the machine-readable format.
A portion is shown here for guidance regarding its form and content.}
\end{deluxetable}

\section{New Occultation Observations} \label{sec:occultations}
\subsection{Spitzer Occultations}

\begin{figure}
\epsscale{1.2}
\plotone{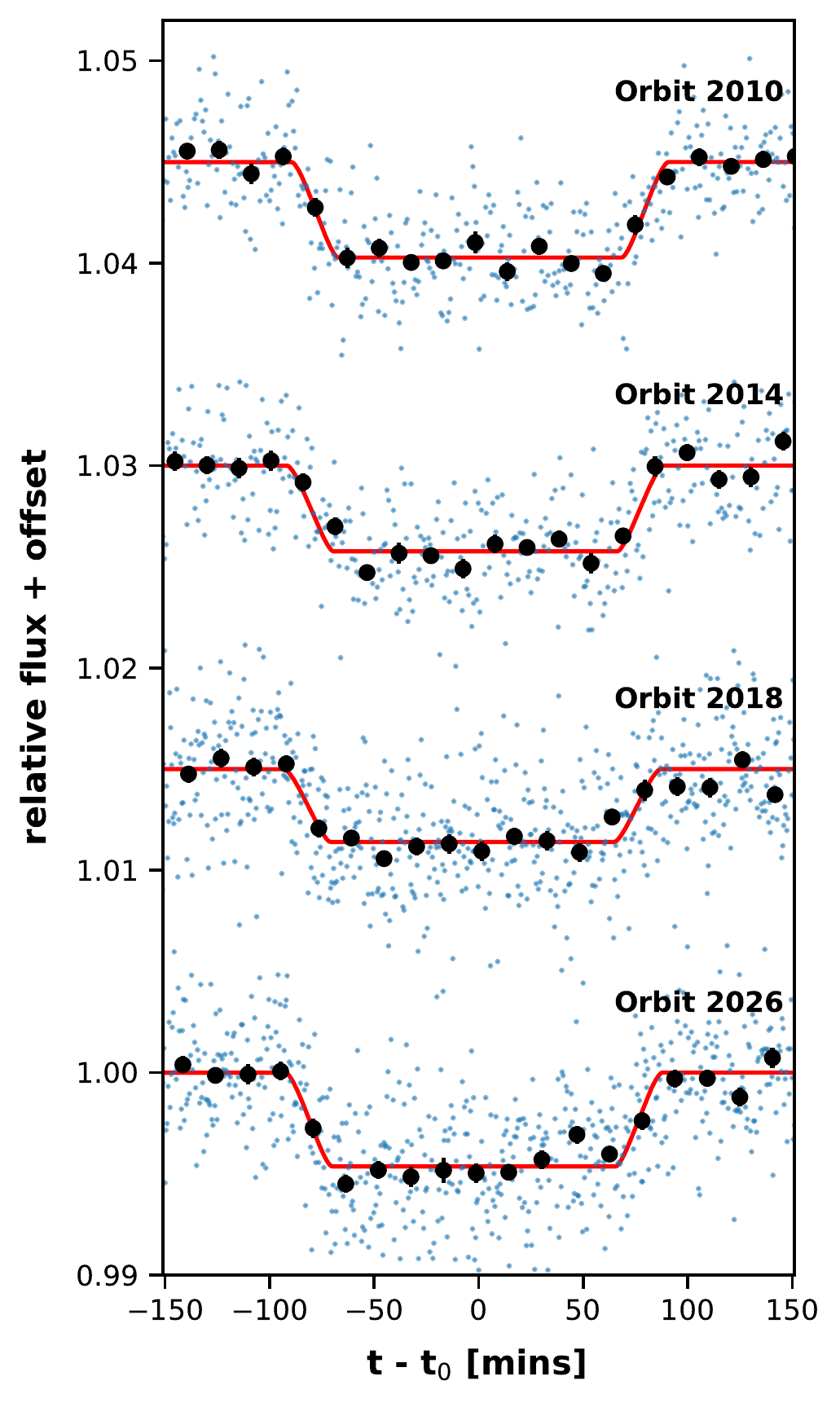}
\caption{Occultations of WASP-12b, based on 4.5\,$\mu$m observations with the \Spitzer Space Telescope.
The data were obtained and analyzed with a time sampling of 2~seconds, but for display purposes
are shown here after averaging in time.
For the top two light curves, the small blue points represent bins of 10 exposures (20 seconds).
For the bottom two light curves, the small blue points represent bins of 14 exposures (28 seconds).
In all cases, the large black points represent bins of 400 exposures (800 seconds).
The red curves are based on the best-fitting model.
As in Figure \ref{fig:transit_lightcurves}, each light curve is labelled with an orbit number relative to a fixed reference orbit.
\label{fig:spitzer_eclipses}}
\end{figure}

\begin{deluxetable*}{lRRRRRRR}
\setlength{\belowrulesep}{0pt}
\tablecolumns{8}
\tablecaption{New Occultation Midpoints and Depths \label{tab:eclipse_fit}}
\tablehead{
	\colhead{\textbf{Source}} &
	\multicolumn{4}{c}{\Spitzer} &
	\colhead{WIRC} &
	\colhead{H19\tablenotemark{a}} &
	\colhead{V19\tablenotemark{b}} \\
	\cmidrule(rl){1-1} \cmidrule(rl){2-5} \cmidrule(rl){6-6} \cmidrule(rl){7-7} \cmidrule(rl){8-8}
	\colhead{\textbf{UT Date}} &
	2019\text{-}01\text{-}16 & 2019\text{-}01\text{-}20 &
	2019\text{-}01\text{-}24 & 2019\text{-}02\text{-}02 &
	2017\text{-}03\text{-}18 & 2017 \text{-}01\text{-}16 &
	2019\text{-}01\text{-}02 \\
	\colhead{\textbf{Orbit Number}} &
	\colhead{2010} & \colhead{2014} & \colhead{2018} & \colhead{2026} &
	\colhead{1397} & \colhead{1341} & \colhead{1997}
}
\startdata
Midpoint\tablenotemark{c} & 58499.75572 & 58504.11988 & 58508.48459 & 58517.21641 &
57830.7139 & 57769.5957 & 58485.5642 \\
Timing Uncertainty &	0.00077 & 0.00087 & 0.00091 & 0.00074 & 0.0011 & 0.0014 & 0.0014 \\
Eclipse Depth (ppm) & 4720^{+289}_{-279}  & 4243^{+270}_{-265} & 3601^{+261}_{-262}	& 4632^{+266}_{-258} &
3232 ^{+110}_{-112} & 1089^{+72}_{-71} & 1095^{+175}_{-176} \\
Photometric band & \multicolumn{4}{c}{Spitzer 4.5~$\mu$m} & $K_\mathrm{s}$ & $i'$ & $V$ \\
\hline
Aperture radius (pixels) & 2.3 & 2.4 & 2.1 & 2.0 & \cdots & \cdots & \cdots \\
Bin size (exposures)	 & 22  & 22  & 14  & 14  & \cdots & \cdots & \cdots \\
\enddata
\tablenotetext{a}{Eclipse observed by \cite{Hooton2019}.}
\tablenotetext{b}{Eclipse observed by \cite{vonEssen2019}.}
\tablenotetext{c}{Times given in $\bjdtdb - 2{,}400{,}000$.}
\tablenotetext{d}{Aperture radius and bin sizes reported in this table only for the \Spitzer observations.}
\end{deluxetable*}

We observed four occultations of WASP-12b with the \Spitzer Space Telescope
in January and February 2019.  The first and last event were separated by sixteen planetary orbits.
All of the data were obtained with the 4.5~\um channel,
in 32$\times$32 pixel sub-array mode with 2-second exposures.
For each event, approximately 11{,}000 exposures were obtained
over a timespan of 7 hours bracketing the 3-hour duration of the occultation.

To reduce the data, we first determined the background level in each exposure by calculating the median flux in the image
after excluding the pixels associated with the host star.
We subtracted this background level from each image.
Then, to measure the pixel location of the centroid of the stellar image, we fitted
a two-dimensional Gaussian function to the central 25 pixels of each exposure.
Using these centroid positions, we performed circular-aperture photometry, with trial aperture radii
ranging from 1.6 to 3.5 pixels in 0.1-pixel increments.
We identified a few outliers based on an unusually large deviation
in the centroid time series; specifically, we flagged any exposures
for which the centroid coordinates were more than 5-$\sigma$
away from the median
of the surrounding 10 exposures.
The flux values for the offending exposures were replaced by the median flux value within that same 10-exposure window.
We also removed from consideration a few data points from the orbit 2026 dataset that had
obvious image artifacts.

To correct for the well-known effects of intra-pixel sensitivity variations, we used the Pixel Level Decorrelation (PLD) technique of \cite{Deming2015}.
We selected a grid of pixels surrounding the stellar image and divided each pixel value by the total flux in that exposure. 
The intention of this normalization procedure is to eliminate the information from the astrophysical signal (which affects all the pixels), leaving behind only the changes due to pointing fluctuations and differences in pixel sensitivity.
Following equation (4) of \cite{Deming2015}, we modeled the light curve as a linear combination of the normalized pixel values $\hat{P}_i(t)$,
a time-dependent trend $ft + gt^2$ that accounts for any phase curve variation or long-term instrumental artifacts,
a constant offset $h$, and a geometric eclipse model $E(t)$ with depth $D$:
\begin{align} \label{eq:pld}
F_{\rm calc}(t) = \sum_{i=1}^{N} c_i \hat{P}_i(t) + ft + gt^2 + h + DE(t).
\end{align}
\added{This model could be extended to include cross-terms between the $\hat{P}_i$ terms and the eclipse model, or higher-order terms in the pixel fluxes \citep{Luger2016}, but we chose not to do so,
given the small values of $\sum c_i \hat{P}_i$ ($< 0.01$) and the eclipse depth ($\sim 0.005$).}
For a given set of eclipse parameters, we used the \texttt{batman} code \citep{Kreidberg2015} to calculate $E(t)$.
We used linear regression to solve Equation~\ref{eq:pld} for the coefficients $c_i$, $f$, $g$, and $D$ that provide the best fit to the data $F_{\rm obs}(t)$.

To speed up computations, it is helpful to reduce the data volume by binning the data in time.
\cite{Deming2015} found that the optimized values of the coefficients $c_i$ sometimes depend on the size of the time bins,
which they attributed to time-correlated noise.
They recommended choosing a bin size that minimizes the amplitude of correlated noise on the timescale of the eclipse.
We determined this optimal bin size as follows.
We obtained an initial estimate for the occultation time by fitting the unbinned data with a model in which all of the eclipse parameters (apart from the occultation time) were fixed to the values found by \cite{Collins2017}.
Then, using a fixed eclipse model with this occultation time, we determined the coefficients $c_i, f, g, h, D$ for time-binned light curves, with bin sizes ranging from 2 to 60 exposures (4 to 120 seconds).
In each case, we examined the residuals by binning them and computing the standard deviation.
For uncorrelated noise, the residuals should scale approximately as $N^{-1/2}$ where $N$ is the number of exposures per bin.
We identified the optimal case as the one for which the residuals best match this expectation.
As a further degree of optimization, we repeated this procedure for each choice of aperture radius, and selected the radius that led to the smallest standard deviation of the residuals.
Table \ref{tab:eclipse_fit} gives the optimal set of photometric parameters for each observation\added{, while the light curves are provided in Table \ref{tab:phot_data}}.

After adopting the optimal aperture and bin size for each of the 4 observations,
we jointly fitted all of the \Spitzer data using a single eclipse model. This time, all of
the eclipse parameters were allowed to vary, subject to prior constraints.
We placed Gaussian priors on the orbital inclination $I$, planet-to-star radius ratio \Rp/\Rstar, and orbit-to-star radius ratio $a/\Rstar$, based on the best-fit values and uncertainties reported by \cite{Collins2017}\added{, shown in Table \ref{tab:stellar_params}}.
These parameters are sufficient to describe the loss of light as a function of the
planet's position on the stellar disk. To specify the loss of light as a function of time for each event,
the timescale $R_\star P/a$ must also be specified [see
Equation (19) of \cite{Winn2010}].
We did so by holding $P$ fixed at the value 1.09142~days,
but importantly, we did not require the interval between occultations to be equal to 1.09142~days.
We allowed the occultation midpoints and depths to be freely varying parameters.

\begin{deluxetable}{cRr}
\tablecolumns{3}
\tablecaption{WASP-12 System Parameters based on \Spitzer Occultations\label{tab:stellar_params}}
\tablehead{
	\colhead{Parameter} &
	\colhead{Prior\tablenotemark{a}} &
	\colhead{Best-fit}
}
\startdata
Period (days)	& 1.09142 	& fixed \\
$\Rp/\Rstar$	& \mathcal{G}(0.11785, 0.00054)	& $0.11786 \pm 0.00027$ \\
$a / \Rstar$	& \mathcal{G}(3.039, 0.034)		& $3.036 \pm 0.014$     \\
$I$ (deg)		& \mathcal{G}(83.37, 0.7)		& $83.38 \pm 0.3$		\\
$e$				& 0.0		& fixed \\
$\omega$ (deg)	& 0.0		& fixed \\
\enddata
\tablenotetext{a}{Based on values from \cite{Collins2017}. We used Gaussian priors denoted by $\mathcal{G}(\mu,\sigma)$.}
\end{deluxetable}

Table \ref{tab:eclipse_fit} gives the final fit \replaced{results}{eclipse times and depths, while Table \ref{tab:stellar_params} gives the remaining fit results}.
The timing precision ranged from 1.0 to 1.3~minutes,
which is similar to the results that were achieved by \cite{Deming2015} and \cite{Patra2017} for the same star. 
Figure \ref{fig:spitzer_eclipses} shows all four detrended \Spitzer light curves, along with the best-fitting eclipse model curves.

\subsection{WIRC Observations}

\begin{figure}
\epsscale{1.2}
\plotone{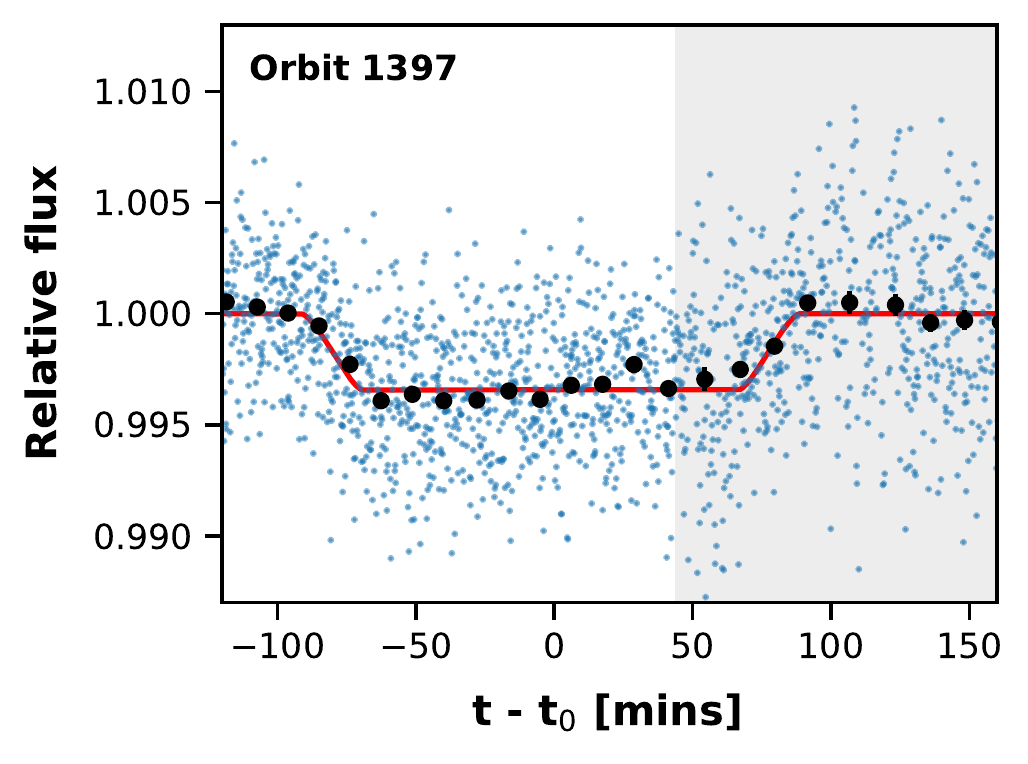}
\caption{Occultation lightcurve observed by WIRC (blue points).
Detrending was performed with the entire lightcurve, but the second half of the lightcurve (shaded gray) was excluded when we performed the final fit to the eclipse model.
The best-fitting eclipse model to the truncated lightcurve is shown in red.
Black points show the lightcurve binned to 70 exposures for clarity.
Error bars are not shown for the unbinned data.
\label{fig:wirc_eclipse}}
\end{figure}

An occultation of WASP-12b was also observed with the Wide-Field Infrared Camera (WIRC, \citealt{Wilson2003a}) on the Hale 200-inch telescope at Palomar Observatory on 18 March 2017.
This observation was made in the $K_\mathrm{s}$ band using a
new Hawaii-II detector installed on WIRC in January 2017 \citep{Tinyanont2019}
and a near-infrared Engineered Diffuser \citep{Stefansson2017}.
We obtained 1{,}828 images with an exposure time of 2 seconds, spanning 5 hours.

Each image was corrected for dark current, flat field, and bad pixels with the WIRC Data Reduction Pipeline \citep{Tinyanont2019}.
We performed circular-aperture photometry of WASP-12 and five comparison stars following the procedure described by \cite{Vissapragada2019}.
A global background was subtracted from each image using iterative 3-$\sigma$
clipping of the flux values, while a local background was determined for each star using annuli with inner and outer radii of 20 and 50 pixels.
Optimizing the pipeline over various aperture sizes found a best circular aperture radius of 9~pixels.
\added{The resulting light curve is provided in Table \ref{tab:phot_data}.}
We modeled the flux time series for WASP-12 as the product of an eclipse model $E(t)$ and
a model of systematic effects, consisting of a linear function of time and 
a linear combination of the fluxes from the five comparison stars:
\begin{align} \label{eq:wirc_model}
M(t) = \left(ft + \sum_{i=1}^{5}c_i S_i(t)\right)\times E(t). 
\end{align}
We used linear regression to solve for the coefficients in the systematics model, given a choice of parameters for the eclipse model \citep{Vissapragada2019}. 

Figure~\ref{fig:wirc_eclipse} shows the resulting light curve after removing the best-fitting model
for the systematic effects. The latter part of the observation was affected by intermittent cirrus clouds,
causing sudden and large-amplitude fluctuations in the measured fluxes
and leading to larger scatter in the detrended light curve.
For this reason, we chose not to fit the data that were obtained during that time period (the gray
region in Figure \ref{fig:wirc_eclipse}). This meant that the egress time and the total transit
duration could not be determined from the WIRC data alone.
Instead, we held fixed the geometric eclipse parameters at the values taken from the best-fitting
model of the \Spitzer data (Table \ref{tab:stellar_params}).
We allowed the eclipse depth and midpoint to vary freely.
The results of this fit are given in Table \ref{tab:eclipse_fit}, and plotted as a
red curve in Figure \ref{fig:wirc_eclipse}.
When the entire light curve is fitted, instead of masking out the latter part of the transit,
the derived mid-eclipse time shifts by 0.5-$\sigma$ and has a formal uncertainty that is a factor of 2 smaller.

\subsection{Re-analysis of Previous Data}

Two other groups have recently reported on observations of occultations of WASP-12b.
\cite{Hooton2019} detected the occultation at the 7-$\sigma$ level using the Isaac Newton Telescope (INT) on La Palma, in January 2017. 
Separately, \cite{vonEssen2019} observed three occultations of WASP-12b in January 2019 with the 2.5m Nordic Optical Telescope (NOT).
While neither of these authors published the mid-eclipse times of their observations, they kindly provided us the light curves.
For the observations by \cite{vonEssen2019}, only the first occultation was securely detected. We only re-analyzed the
data from this event.
We followed the same detrending procedures that are described in their papers
to re-fit the light curves, holding the eclipse parameters fixed at the values
from the best-fitting model to the \Spitzer parameters data (apart from the eclipse depth and midpoint).
We were able to reproduce their results for the eclipse depths.
Table~\ref{tab:eclipse_fit} gives the corresponding mid-eclipse times.

\section{Radial-Velocity Observations} \label{sec:rvs}

\cite{Knutson2014} presented radial-velocity measurements of WASP-12 spanning about 6 years,
using the High Resolution Echelle Spectrometer (HIRES; \citealt{Vogt94}) on the Keck I telescope.
As part of this long-term program, we have obtained three new observations of WASP-12 extending the 
time baseline by 5 years.
These new observations were reduced with the standard pipeline of the California Planet Search (CPS; \citealt{Howard2010}).
Table \ref{tab:rv_data} gives the complete set of radial-velocity data.
The longer baseline is important for detecting any acceleration of the WASP-12 system along the
line of sight, which would lead to apparent changes in orbital period due to the R{\o}mer effect.

\begin{deluxetable}{lrr}
\tablecolumns{3}
\tablewidth{\columnwidth}
\tablecaption{HIRES Radial Velocity Measurements \label{tab:rv_data}}
\tablehead{
	\colhead{Time} &
	\colhead{RV} &
	\colhead{$\sigma(\mathrm{RV})$} \\
	\colhead{\bjdtdb} & \colhead{\ms} & \colhead{\ms}
}
\startdata
2455521.959432 & -136.635 & 2.534 \\
2455543.089922 & 5.728 & 2.919 \\
2455545.983884 & -162.390 & 2.822 \\
2455559.906718 & 141.616 & 2.345 \\
2455559.917563 & 115.818 & 2.727 \\
2455559.927852 & 111.001 & 3.186 \\
2455636.843302 & -143.932 & 2.627 \\
2455671.769904 & -107.997 & 2.446 \\

\enddata
\tablecomments{Table \ref{tab:rv_data} is published in its entirety in the machine-readable format.
A portion is shown here for guidance regarding its form and content.}
\end{deluxetable}

\section{Analysis} \label{sec:analysis}

\begin{deluxetable}{lLLRl}
\tablecolumns{5}
\tablewidth{\columnwidth}
\tablecaption{WASP-12b Transit and Occultation Times \label{tab:timing_data}}
\tablehead{
	\colhead{Event} &
	\colhead{Midtime} &
	\colhead{Error} &
	\colhead{Orbit No.} &
	\colhead{Source} \\
	\colhead{} &
	\colhead{\bjdtdb} &
	\colhead{days} &
	\colhead{} & \colhead{}
}
\startdata
tra & 2454515.52496 & 0.00043 & -1640 & H09$^a$ \\
occ & 2454769.28190 & 0.00080 & -1408 & Ca11 \\
occ & 2454773.64810 & 0.00060 & -1404 & Ca11 \\
tra & 2454836.40340 & 0.00028 & -1346 & C13 \\
tra & 2454840.76893 & 0.00062 & -1342 & Ch11 \\
tra & 2455140.90981 & 0.00042 & -1067 & C17 \\
tra & 2455147.45861 & 0.00043 & -1061 & M13 \\
tra & 2455163.83061 & 0.00032 & -1046 & C17 \\

\enddata
\tablerefs{H09 - \cite{Hebb2009}; C13 - \cite{Copperwheat2013}; C15 - \cite{Croll2015}; C17 - \cite{Collins2017}; Ca11 - \cite{Campo2011}; Ch11 - \cite{Chan2011}; Co12 - \cite{Cowan2012}; Cr12 - \cite{Crossfield2012}; D15 - \cite{Deming2015}; F13 - \cite{Fohring2013}; H19 - \cite{Hooton2019}; K15 - \cite{Kreidberg2015a}; M13 - \cite{Maciejewski2013}; M16 - \cite{Maciejewski2016}; M18 - \cite{Maciejewski2018}; O19 - \cite{Ozturk2019}; P17 - \cite{Patra2017}; P19 - Patra et al. (2019, submitted); S12 - \cite{Sada2012}; S14 - \cite{Stevenson2014}; V19 - \cite{vonEssen2019}}
\tablenotetext{a}{The transit of orbit $-1640$ observed by H09 was reanalyzed by M13.}
\tablenotetext{b}{The occultation of orbit $-722$ observed by D15 was reanalyzed by P17.}
\tablecomments{Table \ref{tab:timing_data} is published in its entirety in the machine-readable format.
A portion is shown here for guidance regarding its form and content.}
\end{deluxetable}

We compiled all of the available transit and occultation times, including the new
data presented in Sections \ref{sec:transits} and \ref{sec:occultations} as well
as from the literature.
We decided to include only those times that were based on fitting the data from
a single event (as opposed to fitting multiple events and requiring periodicity),
for which the midpoint was allowed to be a free parameter, and for
which the time system of the measurement was clearly documented.
Most of these times had already been compiled by \cite{Patra2017};
we added 38 new transit times and 7 new occultation times.
Table \ref{tab:timing_data} gives all of the timing data. We emphasize that
the times in Table \ref{tab:timing_data}
are all in the \bjdtdb system, and that {\it no} adjustment was
made to the observed occultation times to account for the light-travel time across
the diameter of the WASP-12 orbit.  When analyzing the data, as described
below, we {\it did} account for the light-travel time by subtracting
$2a/c=22.9$~seconds from the observed occultation times.\footnote{In the process of compiling the data,
we found that Table 1 of \cite{Patra2017} has an error: except for the \Spitzer data
they presented for the first time, all of the reported occultation
times are wrong by an offset of 50.976 seconds, due to a software bug that
arose from confusion over whether
the light-time correction had already been applied.
This error only affected the values printed in the table, and not the timing analysis
or any of the results reported by \cite{Patra2017}.}

\subsection{Timing Analysis}\label{ssec:timing_analysis}

Following \cite{Patra2017}, we fitted three models to the timing data.
The first model assumes the orbital period to be constant:
\begin{equation}
\begin{aligned}
t_\mathrm{tra}(N) &= t_0 + NP \\
t_\mathrm{occ}(N) &= t_0 + \frac{P}{2} + NP,
\end{aligned}
\end{equation}
where $N$ is the number of orbits from a fixed reference orbit\footnote{We chose the reference orbit as the one with mid-transit time close to $\bjdtdb \approx 2456305.45$, consistent with the choice made in \cite{Patra2017}.},
while $t_0$ is the mid-transit time of this reference orbit.

The second model assumes the orbital period to be changing uniformly with time:
\begin{equation}
\begin{aligned}
t_\mathrm{tra}(N) &= t_0 + NP + \frac{1}{2}\frac{dP}{dN}N^2\\
t_\mathrm{occ}(N) &= t_0 + \frac{P}{2} + NP + \frac{1}{2}\frac{dP}{dN}N^2.
\end{aligned}
\end{equation}

The third model assumes the planet has a nonzero eccentricity $e$ and its
argument of pericenter $\omega$ is precessing uniformly:
\citep{Gimenez1995}:
\begin{equation}
\begin{aligned}
t_\mathrm{tra}(N) &= t_0 + N P_s - \frac{e P_a}{\pi} \cos{\omega(N)}\\
t_\mathrm{occ}(N) &= t_0 + \frac{P_a}{2} + N P_s + \frac{e P_a}{\pi} \cos{\omega(N)}\\
\omega(N) &= \omega_0 + \frac{d\omega}{dN} N \\
P_s &= P_a \left(1 - \frac{1}{2\pi}\frac{d\omega}{dN}\right),
\end{aligned}
\end{equation}
where $P_s$ is the sidereal period and $P_a$ is the anomalistic period. 

\begin{deluxetable}{ll}
\tablecolumns{2}
\tablewidth{\columnwidth}
\tablecaption{Timing Model Fit Parameters \label{tab:timing_fit}}
\tablehead{
	\colhead{Parameter} &
	\colhead{Value (Uncertainty)}
}
\startdata
\multicolumn{2}{l}{\textbf{Constant Period Model}} \\
Period, $P$ (days) 		& \val{constant-period}{per}(\unc{constant-period}{per}) \\
Mid-Transit Time of Reference Orbit, $t_0$	& \val{constant-period}{t0}(\unc{constant-period}{t0}) \\
\hline
N$_\mathrm{dof}$		& \constantperiodmodel{ndof} \\
$\chisqmin$ 			& \constantperiodmodel{chisq} \\
BIC						& \constantperiodmodel{bic}   \\
\hline\hline
\multicolumn{2}{l}{\textbf{Orbital Decay Model}} \\
Period, $P$ (days) 		& \val{decay}{per}(\unc{decay}{per}) \\
Mid-Transit Time of Reference Orbit, $t_0$	& \val{decay}{t0}(\unc{decay}{t0}) \\
Decay Rate, $dP/dN$ (days/orbit)	& \val{decay}{dpde}(\unc{decay}{dpde})$\times10^{-10}$ \\
\hline
N$_\mathrm{dof}$		& \decaymodel{ndof} \\
$\chisqmin$ 			& \decaymodel{chisq} \\
BIC						& \decaymodel{bic}   \\
\hline\hline
\multicolumn{2}{l}{\textbf{Apsidal Precession Model}} \\
Sidereal Period, $P_s$ (days) 		& \val{precess}{per}(\unc{precess}{per}) \\
Mid-Transit Time of Reference Orbit, $t_0$	& \val{precess}{t0}(\unc{precess}{t0}) \\
Eccentricity, $e$		& \val{precess}{e}(\unc{precess}{e}) 	\\
Argument of Periastron, $\omega_0$ (rad)		& \val{precess}{w}(\unc{precess}{w}) 	\\
Precession Rate, $d\omega/dN$ (rad/orbit)	& \val{precess}{wdot}(\unc{precess}{wdot}) 	\\
\hline
N$_\mathrm{dof}$		& \precessmodel{ndof} \\
$\chisqmin$ 			& \precessmodel{chisq} \\
BIC						& \precessmodel{bic}   \\
\hline\hline
\enddata
\tablecomments{Uncertainties in parentheses are the 1-$\sigma$ confidence intervals in the last two digits.}
\end{deluxetable}

\begin{figure*}
\epsscale{1.1}
\plotone{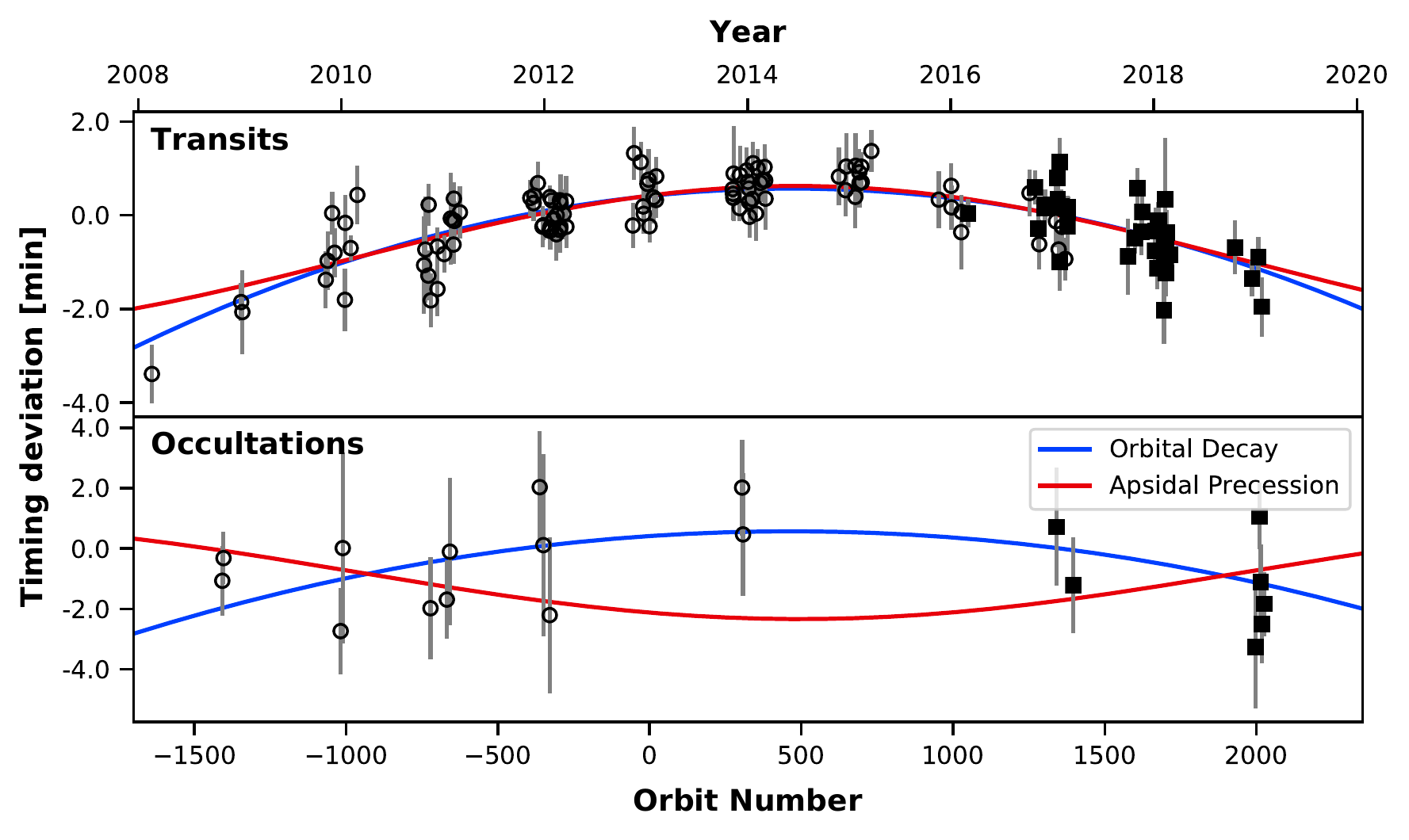}
\caption{Transit and occultation timing residuals, after subtracting the best-fitting constant-period model.
Open circles denote those points previously compiled in \cite{Patra2017}; solid squares are the new transit and occultation times compiled in this work.
The blue line shows the expected residuals for the best-fitting orbital decay model, while the red line shows the best-fitting apsidal precession model.
\label{fig:timing_residuals}}
\end{figure*}

In all 3 cases, we found the best-fitting model parameters by minimizing $\chi^2$.
We again used the \texttt{emcee} code to perform an MCMC sampling of the posterior distribution in parameter space, given broad uniform priors on all the parameters.
We ran the MCMC with 100 walkers, discarding the first \replaced{100,000 steps}{$40\%$ of the steps} as burn-in and \deleted{ensuring convergence by} running the code for $> 10$ autocorrelation times.
\added{We also double-checked convergence by inspecting the posteriors and computing the Geweke scores for each chain \citep{Geweke1992}.}
Table \ref{tab:timing_fit} gives the fit results.

As was already shown by \cite{Maciejewski2016} and \cite{Patra2017}, the constant period model
does not fit the data. The minimum value of $\chi^2$ is \rounded{constant-period}{chisq} with
156 degrees of freedom.
Figure \ref{fig:timing_residuals} shows the residuals.  Also plotted are the best-fitting
model curves for the orbital decay and apsidal precession models.
The best-fitting orbital decay model has $\chisqmin =$ \rounded{decay}{chisq}, while the
best-fitting apsidal precession model has $\chisqmin =$ \rounded{precess}{chisq}.
Thus, while both models fit the data much better than the constant-period model, the orbital
decay model is preferred.  The difference in $\chi^2$ is 12.1.
\cite{Patra2017} also found a preference for orbital decay, but with a weaker statistical
significance ($\dchisq = 5.5$).  Most of the increase in \dchisq is from
the newest \Spitzer observations of occultations, for which the midpoints are consistent
with the predictions of the orbital decay model, but occurred earlier
than would be expected based on the apsidal precession model.

Our confidence that orbital decay is a better description of the data is
enhanced by the fact that the orbital decay model has only 3 free parameters while
the apsidal precession model has 5 free parameters.  A commonly used way to
reward a model for fitting the data with fewer free parameters is to
compare models with the Bayesian Information Criterion (BIC; \citealt{Schwarz1978}):
\begin{equation}
\mathrm{BIC} = \chi^2 + k \log{n},
\end{equation}
where $k$ is the number of free parameters, and $n$ the number of data points.
In this case, the BIC favors the orbital decay model by $\Delta(\mathrm{BIC}) =$ \timingcomparison{dbic}.
The interpretation of this number is not completely straightforward,
but if we assume the posterior distribution of all the parameters to be a multivariate
Gaussian function, then there is a simple
relation between $\Delta$BIC and the Bayes factor $B$:
\begin{equation}
B = \exp{\left[-\Delta(\mathrm{BIC})/2\right]} = 70{,}000,
\end{equation}
representing an overwhelming preference for the orbital decay model.

\subsection{Radial Velocity Analysis} \label{ssec:rv_analysis}
\subsubsection{R{\o}mer Effect} \label{ssec:romer}

\begin{figure}
\epsscale{1.2}
\plotone{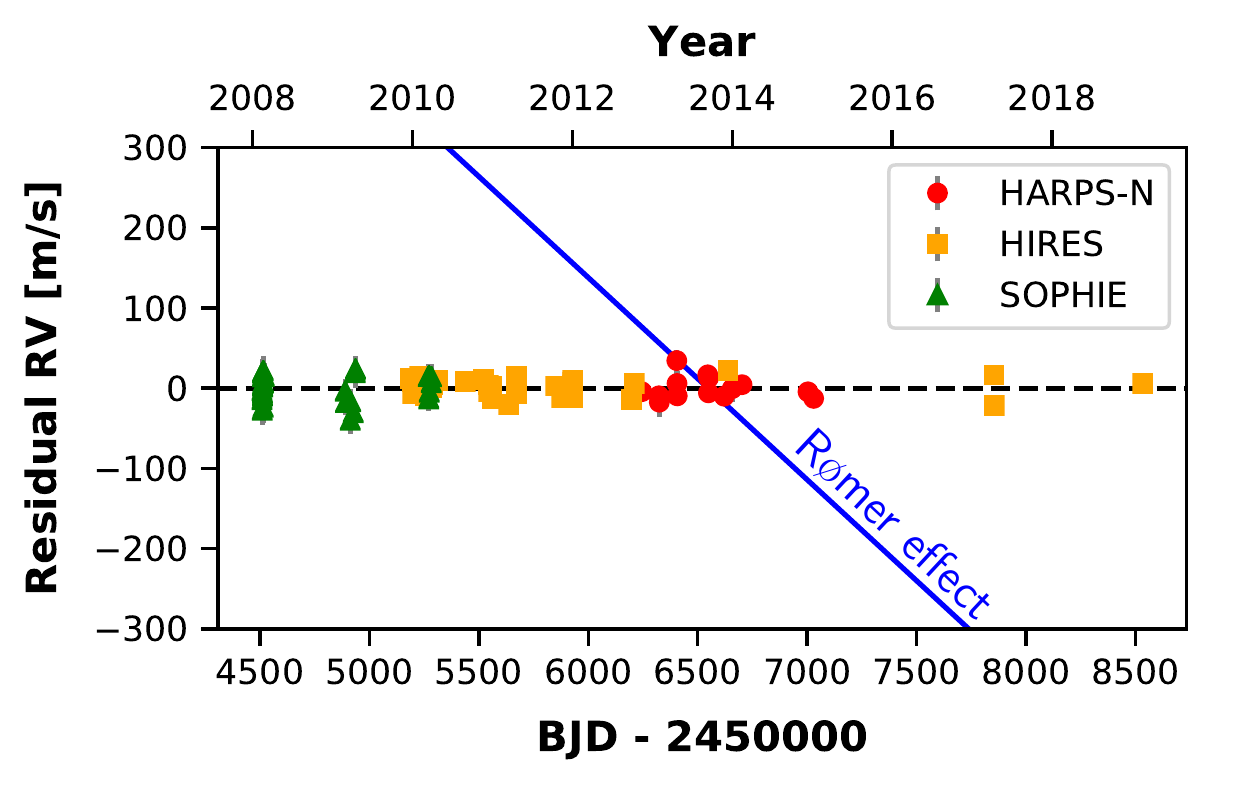}
\caption{Radial velocity residuals after subtracting the best-fitting sinusoidal model.
No significant trends are seen in the residuals.
The blue line has the slope that would have been observed, if the
R{\o}mer effect were solely responsible for the observed period derivative.
\label{fig:rv_residuals}}
\end{figure}

If the center of mass of the
star-planet system is accelerating along the line of sight
with a magnitude $\dot{v}_r$,
then the apparent period of the hot Jupiter would be observed to change
due to the R{\o}mer effect:
\begin{equation*}
\frac{\dot{P}}{P} = \frac{\dot{v}_r}{c}.
\end{equation*}
If we insert the measured values of $P$ and $\dot{P}$ for WASP-12 into
this equation, the implied acceleration
is $\dot{v}_r = 0.25$~m~s$^{-1}$~day$^{-1}$.
This is more than an order of magnitude larger than the
2-$\sigma$ upper limit of 0.019~m~s$^{-1}$~day$^{-1}$ that
\cite{Patra2017} obtained by fitting the
previously available radial-velocity data.
Here, we use the newly obtained radial-velocity data to place an even more stringent upper limit.

In addition to the HIRES radial velocities from \cite{Knutson2014} and described in Section \ref{sec:rvs}, we analyzed the data obtained with the SOPHIE spectrography by \cite{Hebb2009} and \cite{Husnoo2011}, as well as data obtained with the HARPS-N spectrograph by \cite{Bonomo2017}. We allowed for a constant
velocity offset and a ``jitter'' value specific to each spectrograph. We also excluded the data points
obtained during transits, to avoid having to model the Rossiter-McLaughlin effect.
We used the \texttt{radvel} code \citep{Fulton18} to fit a model consisting of the sum of a sinusoidal
function (representing the circular orbit of the planet) and a linear function of time (representing a constant
radial acceleration). We fixed the period and time of conjunction to the values
from the best-fitting constant period timing model (Table \ref{tab:timing_fit}).

Figure \ref{fig:rv_residuals} shows the residuals, after subtracting the best-fitting model.
Any line-of-sight acceleration must have an amplitude $|\dot{v_r}| < 0.005$~\msday, at the 2-$\sigma$ level.
This is a factor of four improvement over the constraints reported by \cite{Patra2017},
and two orders of magnitude smaller than the value that would be observed if the observed
period derivative were entirely due to the R{\o}mer effect.
Thus, we can dismiss the R{\o}mer effects as a significant contributor to the observed period change.
\footnote{\added{Based on a re-analysis of the SOPHIE and HARPS spectra, \cite{Baluev2019} reported a radial-velocity trend of $-7.5\pm 2.2$~\msyr or $-0.021\pm 0.006$~\msday.
This is 4 times larger than the than our upper limit, and is therefore ruled out. 
The most recent Keck/HIRES data were helpful in this regard.}}

\subsubsection{Orbital Eccentricity} \label{ssec:eccentricity}

The apsidal precession hypothesis requires that the orbit is slightly eccentric.
In the best-fitting timing model, the eccentricity is about $0.003$.  This raises a theoretical problem, because the expected timescale for tidal circularization is about 0.5~Myr \citep{Patra2017}.
Thus, if apsidal precession were taking place, there needs to be some mechanism for
maintaining the eccentricity at the level of $10^{-3}$ despite the expectation of rapid tidal circularization.

For WASP-12b, the radial-velocity data are compatible with a circular orbit, but they
are not precise enough to rule out an orbital
eccentricity at the level of $10^{-3}$.
Indeed, when we refitted the radial-velocity data allowing the orbit to be eccentric,
we found that the best-fitting model implies a {\it larger} eccentricity:
$\secosw =$~\rveccmodel{secosw}, $\sesinw =$~\rveccmodel{sesinw}, and
$e =$~\rveccmodel{e}.  This is consistent with the previous reports of a nonzero eccentricity
by \cite{Knutson2014} and \cite{Husnoo2011}.

However, we think that this is a spurious result. The clue is that the best-fitting
argument of pericenter is aligned nearly exactly with the line of sight: $\omega =$~\rveccmodel{w}.
Most likely, the orbit is circular, and the apparently nonzero eccentricity is due to an unmodeled systematic effect.
A good candidate for this effect is the tidal distortion of the host star.
As pointed out by \cite{Arras2012}, the tidal bulge raised by a hot Jupiter can cause an apparent radial-velocity signal with a period equal to half of the orbital period.
This would lead to a second harmonic in the radial-velocity data that can be mistaken for the signal of an eccentric orbit.
In particular, \cite{Arras2012} showed that a planet on a circular orbit would appear to have a nonzero orbital eccentricity and an argument of pericenter $\omega = -90$~deg.
Furthermore, \cite{Arras2012} predicted that for the specific case of WASP-12, the fictitious eccentricity would have an amplitude on the order of 0.02. 
Both of these predictions are consistent with what has
been observed.\footnote{At the Extreme Solar Systems IV conference, in Reykjavik, Iceland (August 19-23, 2019), G.\ Maciejewski presented further evidence for the effect of the tidal bulge
in the radial-velocity data for WASP-12.}

\subsubsection{Joint RV-Timing Fit} \label{ssec:rv_timing}

In principle, the radial-velocity signal should also be affected
by either orbital decay or apsidal precession.
A change in the orbital period would affect the RV signal via the true anomaly $\nu$.
We computed the radial-velocity signal of a planet with a constant and decaying period using the parameters in Table \ref{tab:timing_fit}.
Over the 9-year timespan of the HIRES radial-velocity observations, we found that the maximum deviation between the two models is $\sim 10\,\ms$.

As for apsidal precession, \cite{Csizmadia2019} presented a formula for the associated RV signal:
\begin{equation}
\begin{aligned}
RV = K & \left[e\cos{\omega(t)} + \cos{\left(\nu+\omega(t)\right)} \right. \\
& \left. + \frac{\dot{\omega}}{n} \frac{\left(1-e^2\right)^{3/2} \cos{\left(\nu+\omega(t)\right)}}{1 + e\cos{\nu}}\right] \\
\omega(t) = \omega_0 & + \dot{\omega}\left(t - t_0\right),
\end{aligned}
\end{equation}
where $n\equiv 2\pi/P$ is the mean motion.
Based on this equation, along with the precession period from the best-fitting apsidal precession model, we found that the maximum deviation between the RV signal of a precessing orbit with parameters in Table \ref{tab:timing_fit} and a circular model would be $\sim 6\,\ms$ over a decade.
We note that \cite{Csizmadia2019} claimed that apsidal precession would result in residuals on the order of $K_\mathrm{orb}$; however that only holds if the anomalistic period could be determined independently.
In reality, the anomalistic period would need to be determined using the same data, and the residuals would be on the order of $e K_\mathrm{orb}$.

In both cases, there should be a small but potentially measurable effects on the RV data.
We tried fitting the timing and RV datasets jointly, but the results
were essentially unchanged. 
In particular, the RV data did not alter the \dbic between the orbital decay and apsidal precession models.
This is likely because the RV observations at later times are sparsely sampled, preventing these small deviations from being measured effectively.
Furthermore, the RV jitter of WASP-12 in the HIRES observations is $\sigjit \sim 10\,\ms$, a similar magnitude to the decay or precession RV signal.
Future RV measurements could be helpful, but only if the RV systematic effects are better understood, including the tidal effect described in Section \ref{ssec:eccentricity}. 

\section{Discussion} \label{sec:discussion}

Since the work of \cite{Patra2017}, evidence has continued to mount that the orbit of
WASP-12b is decaying, as opposed to apsidally precessing.
The difference in $\chi^2$ between the orbital decay and apsidal precession models
has grown from 5.5 to \timingcomparison{dchisq}, and the difference in the Bayesian
Information Criterion has widened from 14.9 to \timingcomparison{dbic}.
Also noteworthy is that as new data has become available,
the best-fitting orbital decay parameters have remained the same,
while the best-fitting apsidal precession parameters have changed significantly.
Our measurement of $dP/dN =$~\decaymodel{dpde} is consistent with the rate of $(-10.2 \pm 1.1)\times10^{-10}$~days/orbit reported by \cite{Patra2017},
and with the rate of $dP/dN = (-9.67 \pm 0.73)\times10^{-10}$~days/orbit
reported
by
\cite{Maciejewski2018}.
In contrast, between 2017 and our study,
the best-fitting orbital eccentricity in the precession model has increased by a factor of $1.5\pm 0.4$ and the best-fitting precession period has increased by a factor of $1.4\pm 0.2$.

We are now ready to conclude that WASP-12b is the first planet known to be undergoing
orbital decay.  
The timescale over which the orbit is shrinking is
\begin{equation*}
\tau = \frac{P}{|\dot{P}|} = 3.25^{+0.24}_{-0.21}\,\mathrm{Myr}.
\end{equation*}
Given that the host star appears to be at least 1~Gyr old,
it may seem remarkable that we are observing the planet so close to the time of its destruction.
If we were observing a single planet at a random moment within its 1~Gyr lifetime,
the chance of catching it within the last 3~Myr would be only 0.3\$.
However, ground and space-based surveys have searched hundreds of thousands of stars
for hot Jupiters.
Given that hot Jupiters occur around $\sim$0.1\% of stars and have a transit probability of $\sim$10\%, we might expect to find $500{,}000 \times 1\% \times 10\% \times 0.3\% \sim 2$ planets as close to the end of their lives as is implied by the decay rate of WASP-12b. 

The orbital energy and angular momentum are both decreasing, at rates of
\begin{align}
    \frac{dE}{dt} &= \frac{(2\pi)^{2/3} M_{\rm p}}{3} \left(\frac{GM_\star}{P} \right)^{2/3} \frac{1}{P}\frac{dP}{dt} = -5 \times 10^{23}~{\rm Watts},\\
    \frac{dL}{dt} &= \frac{M_{\rm p}}{3(2\pi)^{1/3}}  \left(\frac{GM_\star}{P} \right)^{2/3} \frac{dP}{dt} = -7\times 10^{27}~{\rm kg\,m}^{2}\,{\rm s}^{-2},\label{eq:angmom_loss}
\end{align}
\added{where we have used the stellar and planetary masses $M_\star = 1.4\pm0.1\,\Msun, \Mp = 1.47 \pm 0.07\,\Mjup$, from \cite{Collins2017}.}
While it would still be useful to reduce the uncertainties in the timing model by observing more transits and occultations, there are more interesting questions regarding the mechanism by which the orbit is losing energy and angular momentum, and what sets WASP-12b apart from the other hot Jupiters for which orbital decay has not been detected.

\subsection{Tidal Orbital Decay}

The possibility discussed in the Introduction is that the angular momentum is being
transferred directly to the star through the gravitational torque on the star's tidal bulge,
and the energy is being dissipated inside the star as the tidal oscillations
are converted into heat.
In the ``constant phase lag'' model of \cite{Goldreich1966}, assuming that
the planet's mass stays constant, the decay rate is
\begin{equation*}
\dot{P} = -\frac{27\pi}{2Q_*'}\left(\frac{M_p}{M_\star}\right)\left(\frac{R_\star}{a}\right)^5,
\end{equation*}
where $Q'_\star$ is the star's ``modified tidal quality factor'', defined as the quality
factor $Q_\star$ divided by $2/3$ of the Love number $k_2$.
Inserting the measured decay rate, we obtain for WASP-12 a tidal quality factor of
\begin{equation*}
Q'_\star = 1.75^{+0.13}_{-0.11}\times10^{5}.
\end{equation*}

This result for $Q'_\star$ is lower than most of the
estimates in the literature, which are based on less direct
observations.
By analyzing the eccentricity distribution of stellar binaries, \cite{Meibom2005} found $\Qs$ to be in the range
from $10^5$ to $10^7$, while similar studies applied to hot Jupiter systems by \cite{Jackson2008} and \cite{Husnoo2012} found $\Qs = 10^{5.5} - 10^{6.5}$.
\cite{Hamer2019} found that hot Jupiter host stars
are kinematically younger than similar stars without
hot Jupiters, and interpreted the result as evidence
for the tidal destruction of hot Jupiters, finding
\Qs~$= 10^6 - 10^{6.5}$.
\cite{CollierCameron2018} modeled the orbital period distribution of the hot Jupiter population and found $\Qs=10^7-10^8$.
\cite{Penev2012} modeled the star/planet tidal interactions in selected systems and also found $\Qs > 10^7$.
However, these studies assumed
$Qs$ to be a universal constant, even though one would
expect it to depend on forcing frequency and perhaps
many other parameters. \cite{Penev2018}
updated and expanded the approach of
system-by-system modeling to allow for frequency
dependence, finding
that \Qs ranges from $10^5$ to $10^7$ for orbital periods of 0.5 to 2 days \citep{Penev2018}.

If tidal dissipation is responsible for the orbital decay of WASP-12,
then the dissipation rate is higher than would have been expected
according to these earlier studies.
The physical mechanism for such rapid dissipation is unclear.
Attempts to compute the tidal quality factor from physical principles for either the equilibrium or dynamical tide generally find larger values of \Qs, from~$10^7$ to $10^{10}$ (as
reviewed by \citealt{Ogilvie2014}).
\cite{Weinberg2017} argued that WASP-12 could be a subgiant star, in
which case nonlinear wave-breaking of the dynamical tide near the stellar core
would lead efficient dissipation and $\Qs \sim 2\times10^5$.
However, \cite{Bailey2019} examined this possibility and 
found that the observed properties of WASP-12
are more compatible with the expected properties of a main-sequence star
rather than a subgiant.

\cite{Millholland2018} proposed an alternate hypothesis, in which
WASP-12b is in a spin-orbit resonance with an external perturber, allowing
it to maintain a large obliquity and giving rise to
obliquity tides.
The ongoing dissipation of obliquity tides might be strong 
enough to explain the observed decay rate.
While such a hypothetical perturber cannot yet be ruled out, 
this scenario calls for a specific system architecture with a misaligned perturber just below the limits of detection.

\subsection{Roche Lobe Overflow}

The Roche limit for a close-orbiting planet can be expressed as a minimum orbital period depending on the density distribution of the planet \citep{Rappaport2013}.
For WASP-12b, with a mean density of 0.46~g~cm$^{-3}$, the Roche-limiting orbital period is 14.2~hr assuming the mass of the planet to be concentrated near
the center, and 18.6~hr in the opposite limit of a spherical and incompressible planet.
The true orbital period of 26~hr is longer than either of these limits, 
implying that the planet is not filling its Roche lobe.  This
simple comparison does not take into account the planet's tidal distortion, which
would lead to a longer period for the Roche limit, but probably not as long as 26~hr.

Nevertheless, there may exist an optically thin exosphere or wind that does fill the Roche lobe. 
Ultraviolet observations of the WASP-12 system indicate that the planet is surrounded by a cloud of absorbing material that is larger than the Roche lobe \citep{Fossati2010,Haswell2012,Nichols2015}.
Recent infrared phase curve observations of the WASP-12 system also corroborate the idea that gas is being stripped from the planet \citep{Bell2019}.
The mass-loss rate is not measured directly \citep{Haswell2018}, but models of this process suggest it could be as high as $\sim 3\times10^{14}$~g~s$^{-1}$, corresponding to $M_\mathrm{p}/\dot{M} \approx 300$~Myr \citep{Lai2010,Jackson2017}.
This mass-loss timescale, while longer than the tidal decay timescale,
is still short relative to the age of the star.
Putting this evidence together with the observed changes in orbital period,
it seems that tidal orbital decay has brought the planet close enough to initiate Roche
lobe overflow of the planet's tenuous outer atmosphere.

The escaping mass bears energy and angular momentum, which can lead to changes in the orbital period separate from the effects of the tidal bulge of the star.
To assess whether or not the escaping mass bears enough angular momentum to be relevant for period changes, we need to know more about the flow of mass away from the planet.
We would expect the gas to flow out from the inner Lagrange point, with lower specific angular momentum than the rest of the planet.
\added{
As a result, the planet's specific angular momentum would rise, causing its orbit to widen.
In this scenario, the rate of period decrease that we have measured would be the result of a competition between tidally-driven decay and mass-loss-driven growth.
\cite{Jia2017} presented an expression relating the change in semimajor axis to the mass loss rate:
\begin{equation}
\frac{\dot{a}}{a} = \frac{2}{1+q}\frac{\dot{M}_\mathrm{p}}{M_\mathrm{p}}\left[\frac{x_\mathrm{L1}^2}{x_\mathrm{p}^2}(1-\epsilon) + q^2 - 1\right] > 0,
\end{equation}
where $q$ is the planet-star mass ratio, $x_\mathrm{L1}, x_\mathrm{p}$ are the distances from the L1 point and planet to the system center of mass, and $\epsilon$ parameterizes the fraction of angular momentum that is transferred back to the planet from the outflowing gas.
Using the mass-loss rate estimated by \cite{Lai2010} and \cite{Jackson2017}, and assuming no angular momentum transfer back to the planet ($\epsilon = 0$), this yields
\begin{align}
\frac{\dot{P}}{P} &\approx 3\frac{\dot{M}_\mathrm{p}}{M_\mathrm{p}} \\
\dot{P} &\approx 1~\mathrm{msec}\,\mathrm{yr}^{-1}.
\end{align}
Under these assumptions, the mass loss from the planet causes a period increase about 30 times smaller than the observed period change for WASP-12b, and is unlikely to be slowing down the planet's orbital decay.

On the other hand, if gas flows out of the L2 point, there could be a net loss in angular momentum from the planet, hastening any tidally-driven decay of the orbit.
\cite{Valsecchi2015} examined this process and found that
the orbital period begins to shrink only in the final stages of mass loss, when
the remaining mass of the planet is dominated by the dense core.
This is not the current situation of WASP-12b, given its Jovian mass and low mean density.
Furthermore, \cite{Jia2017} found that even in such a scenario, the mass loss from L1 continues to dominate over the outflow from L2, leading to continued expansion of the orbit.
In either case, it seems unlikely that the mass loss from WASP-12b is  contributing significantly to its orbital evolution.
}

\deleted{
This would cause the planet's specific angular momentum to rise, and the orbit to widen, rather than shrink.
In this scenario, the rate of period decrease that we have measured would be the result of a competition between tidally-driven decay and mass-loss-driven growth.
However, if gas flows out of the L2 point, there could be a net loss in angular momentum from the planet, leading to decay of the orbit.
\cite{Valsecchi2015} examined this process and found that
the orbital period begins to shrink only in the final stages of mass loss, when
the remaining mass is dominated by the dense core of the planet.
This is not the current situation of WASP-12b, given its Jovian mass
and low mean density.
}

\deleted{
Furthermore, if we compare the rate of angular momentum loss $dL/dt$ to the specific angular momentum at the planet's orbital distance, $l = \sqrt{GM_\star a}$, we find
This is the implied mass loss rate assuming all the gas were lost with specific angular momentum equal to the orbital angular momentum. It is about 30 times larger than the mass loss rate estimated by \cite{Lai2010} and \cite{Jackson2017}.
Thus, it seems that the mass loss from WASP-12b is unlikely to be affecting
its orbital evolution, and tidal dissipation within the star is probably the dominant mechanism responsible for the orbital decay.
}

\section{Conclusion}

We have presented new timing data for WASP-12b that have finally allowed us to determine that its orbit is decaying.
We measured a shift in transit and occultation times of about 4~minutes over a ten-year period,
corresponding to a period derivative of $\dot{P} = $~\decaymodel{pdot_yr}.
This is the first time a hot Jupiter has been caught in the act of spiraling into its host star. It
will likely be destroyed on a timescale of several million years.

The WASP-12 system will be observed by the \textit{Transiting Exoplanet Survey Satellite} (\TESS; \citealt{Ricker14}) in Sector 20, from December 2019 to January 2020.
The new high-precision light curves from \TESS will improve our measurement of \Qs of WASP-12, as will any further transit or occultation observations in the future.

\added{It will also be important to seek evidence for orbital decay in other systems. Already, it is clear that not all systems have the same effective value of $Q_\star'$.
For example, the hot Jupiter WASP-19b has an orbital period of 0.79 days, even shorter than that of WASP-12,
and was predicted to be the most favorable system for measuring orbital decay (see, e.g. \citealt{Essick2015,Valsecchi2014b}).
However, a recent analysis of transit times by \cite{Petrucci2019} shows that any period changes of WASP-19b are slower than 2.2~msec~yr$^{-1}$, implying $Q'_\star<1.2\times 10^6$, limits that are incompatible with the observations of WASP-12.
This may be because WASP-12 is a subgiant star, as advocated by \cite{Weinberg2017}.
The best way to understand if this is the case is to expand the collection of systems
for which detections or period changes, or stringent upper limits, have been made.
This will help to clarify the interpretation and check on the dependence of
tidal dissipation rates on properties such as the planet mass and orbital period, and the
stellar evolutionary state and rotation rate.}

\added{In the case of WASP-12b,} it was only through more than a decade of combined photometric and spectroscopic observations that we were able to firmly distinguish the orbital decay scenario from other physical processes that can result in similar observational signatures.
Similar monitoring of other hot Jupiters - and indeed, all types of planets - will be important in understanding these slow-acting but important physical processes which sculpt the architecture of exoplanetary systems.

\acknowledgements
We thank Jeremy Goodman and Luke Bouma for helpful discussions while preparing this manuscript.
The authors gratefully acknowledge Leo Liu for helping obtain the WIRC observations.
Work by S.W.Y and J.N.W was partly supported by the Heising-Simons Foundation.
J.T.W. acknowledges support from NASA Origins of Solar Systems grant NNX14AD22G. The Center for Exoplanets and Habitable Worlds is supported by the Pennsylvania State University, the Eberly College of Science, and the Pennsylvania Space Grant Consortium.
This research has made use of the VizieR catalogue access tool, CDS, Strasbourg, France.
The original description of the VizieR service was published in A\&AS 143, 23.
Some of the data presented herein were obtained at the W. M. Keck Observatory, which is operated as a scientific partnership among the California Institute of Technology, the University of California and the National Aeronautics and Space Administration. The Observatory was made possible by the generous financial support of the W. M. Keck Foundation.
The authors wish to recognize and acknowledge the very significant cultural role and reverence that the summit of Maunakea has always had within the indigenous Hawaiian community.  We are most fortunate to have the opportunity to conduct observations from this mountain.

\software{\texttt{astropy} \citep{Astropy13,Astropy18};
\texttt{emcee} \citep{Foreman-Mackey13,Goodman10};
\texttt{radvel} \citep{Fulton18};
\texttt{numpy} \citep{Numpy};
\texttt{scipy} \citep{Scipy}}

\clearpage

\bibliography{manuscript,software,instruments}{}

\begin{thebibliography}{}
\expandafter\ifx\csname natexlab\endcsname\relax\def\natexlab#1{#1}\fi
\providecommand{\url}[1]{\href{#1}{#1}}
\providecommand{\dodoi}[1]{doi:~\href{http://doi.org/#1}{\nolinkurl{#1}}}
\providecommand{\doeprint}[1]{\href{http://ascl.net/#1}{\nolinkurl{http://ascl.net/#1}}}
\providecommand{\doarXiv}[1]{\href{https://arxiv.org/abs/#1}{\nolinkurl{https://arxiv.org/abs/#1}}}

\bibitem[{Arras {et~al.}(2012)Arras, Burkart, Quataert, \&
  Weinberg}]{Arras2012}
Arras, P., Burkart, J., Quataert, E., \& Weinberg, N.~N. 2012, Monthly Notices
  of the Royal Astronomical Society, 422, 1761,
  \dodoi{10.1111/j.1365-2966.2012.20756.x}

\bibitem[{{Astropy Collaboration} {et~al.}(2013){Astropy Collaboration},
  {Robitaille}, {Tollerud}, {Greenfield}, {Droettboom}, {Bray}, {Aldcroft},
  {Davis}, {Ginsburg}, {Price-Whelan}, {Kerzendorf}, {Conley}, {Crighton},
  {Barbary}, {Muna}, {Ferguson}, {Grollier}, {Parikh}, {Nair}, {Unther},
  {Deil}, {Woillez}, {Conseil}, {Kramer}, {Turner}, {Singer}, {Fox}, {Weaver},
  {Zabalza}, {Edwards}, {Azalee Bostroem}, {Burke}, {Casey}, {Crawford},
  {Dencheva}, {Ely}, {Jenness}, {Labrie}, {Lim}, {Pierfederici}, {Pontzen},
  {Ptak}, {Refsdal}, {Servillat}, \& {Streicher}}]{Astropy13}
{Astropy Collaboration}, {Robitaille}, T.~P., {Tollerud}, E.~J., {et~al.} 2013,
  \aap, 558, A33, \dodoi{10.1051/0004-6361/201322068}

\bibitem[{{Astropy Collaboration} {et~al.}(2018){Astropy Collaboration},
  {Price-Whelan}, {Sip{\H{o}}cz}, {G{\"u}nther}, {Lim}, {Crawford}, {Conseil},
  {Shupe}, {Craig}, {Dencheva}, {Ginsburg}, {Vand erPlas}, {Bradley},
  {P{\'e}rez-Su{\'a}rez}, {de Val-Borro}, {Aldcroft}, {Cruz}, {Robitaille},
  {Tollerud}, {Ardelean}, {Babej}, {Bach}, {Bachetti}, {Bakanov}, {Bamford},
  {Barentsen}, {Barmby}, {Baumbach}, {Berry}, {Biscani}, {Boquien}, {Bostroem},
  {Bouma}, {Brammer}, {Bray}, {Breytenbach}, {Buddelmeijer}, {Burke},
  {Calderone}, {Cano Rodr{\'\i}guez}, {Cara}, {Cardoso}, {Cheedella}, {Copin},
  {Corrales}, {Crichton}, {D'Avella}, {Deil}, {Depagne}, {Dietrich}, {Donath},
  {Droettboom}, {Earl}, {Erben}, {Fabbro}, {Ferreira}, {Finethy}, {Fox},
  {Garrison}, {Gibbons}, {Goldstein}, {Gommers}, {Greco}, {Greenfield},
  {Groener}, {Grollier}, {Hagen}, {Hirst}, {Homeier}, {Horton}, {Hosseinzadeh},
  {Hu}, {Hunkeler}, {Ivezi{\'c}}, {Jain}, {Jenness}, {Kanarek}, {Kendrew},
  {Kern}, {Kerzendorf}, {Khvalko}, {King}, {Kirkby}, {Kulkarni}, {Kumar},
  {Lee}, {Lenz}, {Littlefair}, {Ma}, {Macleod}, {Mastropietro}, {McCully},
  {Montagnac}, {Morris}, {Mueller}, {Mumford}, {Muna}, {Murphy}, {Nelson},
  {Nguyen}, {Ninan}, {N{\"o}the}, {Ogaz}, {Oh}, {Parejko}, {Parley}, {Pascual},
  {Patil}, {Patil}, {Plunkett}, {Prochaska}, {Rastogi}, {Reddy Janga},
  {Sabater}, {Sakurikar}, {Seifert}, {Sherbert}, {Sherwood-Taylor}, {Shih},
  {Sick}, {Silbiger}, {Singanamalla}, {Singer}, {Sladen}, {Sooley},
  {Sornarajah}, {Streicher}, {Teuben}, {Thomas}, {Tremblay}, {Turner},
  {Terr{\'o}n}, {van Kerkwijk}, {de la Vega}, {Watkins}, {Weaver}, {Whitmore},
  {Woillez}, {Zabalza}, \& {Astropy Contributors}}]{Astropy18}
{Astropy Collaboration}, {Price-Whelan}, A.~M., {Sip{\H{o}}cz}, B.~M., {et~al.}
  2018, \aj, 156, 123, \dodoi{10.3847/1538-3881/aabc4f}

\bibitem[{Bailey \& Goodman(2019)}]{Bailey2019}
Bailey, A., \& Goodman, J. 2019, Monthly Notices of the Royal Astronomical
  Society, 482, 1872, \dodoi{10.1093/mnras/sty2805}

\bibitem[{Baluev {et~al.}(2019)Baluev, Sokov, Jones, Shaidulin, Sokova,
  Nielsen, Benni, Schneiter, D'Angelo, {Fern{\'a}ndez-Laj{\'u}s}, Di~Sisto,
  Ba{\c s}t{\"u}rk, Bretton, Wunsche, Hentunen, Shadick, Jongen, Kang, Kim,
  Pak{\v s}tien{\.e}, Qvam, Knight, Guerra, Marchini, Salvaggio, Papini, Evans,
  Salisbury, Garcia, Molina, Garlitz, Esseiva, Ogmen, Karavaev, Rusov,
  Ibrahimov, \& Karimov}]{Baluev2019}
Baluev, R.~V., Sokov, E.~N., Jones, H. R.~A., {et~al.} 2019, arXiv:1908.04505
  [astro-ph].
\newblock \doarXiv{1908.04505}

\bibitem[{Bell {et~al.}(2019)Bell, Zhang, Cubillos, Dang, Fossati, Todorov,
  Cowan, Deming, Zellem, Stevenson, Crossfield, {Dobbs-Dixon}, Fortney,
  Knutson, \& Line}]{Bell2019}
Bell, T.~J., Zhang, M., Cubillos, P.~E., {et~al.} 2019, Monthly Notices of the
  Royal Astronomical Society, 489, 1995, \dodoi{10.1093/mnras/stz2018}

\bibitem[{Bonomo {et~al.}(2017)Bonomo, Desidera, Benatti, Borsa, Crespi,
  Damasso, Lanza, Sozzetti, Lodato, Marzari, Boccato, Claudi, Cosentino,
  Covino, Gratton, Maggio, Micela, Molinari, Pagano, Piotto, Poretti,
  Smareglia, Affer, Biazzo, Bignamini, Esposito, Giacobbe, H{\'e}brard,
  Malavolta, Maldonado, Mancini, Fiorenzano, Masiero, Nascimbeni, Pedani,
  Rainer, \& Scandariato}]{Bonomo2017}
Bonomo, A.~S., Desidera, S., Benatti, S., {et~al.} 2017, Astronomy \&
  Astrophysics, 602, A107, \dodoi{10.1051/0004-6361/201629882}

\bibitem[{Campo {et~al.}(2011)Campo, Harrington, Hardy, Stevenson, Nymeyer,
  Ragozzine, Lust, Anderson, {Collier-Cameron}, Blecic, Britt, Bowman,
  Wheatley, Loredo, Deming, Hebb, Hellier, Maxted, Pollaco, \&
  West}]{Campo2011}
Campo, C.~J., Harrington, J., Hardy, R.~A., {et~al.} 2011, The Astrophysical
  Journal, 727, 125, \dodoi{10.1088/0004-637X/727/2/125}

\bibitem[{Chan {et~al.}(2011)Chan, Ingemyr, Winn, Holman, {Sanchis-Ojeda},
  Esquerdo, \& Everett}]{Chan2011}
Chan, T., Ingemyr, M., Winn, J.~N., {et~al.} 2011, The Astronomical Journal,
  141, 179, \dodoi{10.1088/0004-6256/141/6/179}

\bibitem[{Claret \& Bloemen(2011)}]{Claret2011}
Claret, A., \& Bloemen, S. 2011, Astronomy \& Astrophysics, 529, A75,
  \dodoi{10.1051/0004-6361/201116451}

\bibitem[{{Collier Cameron} \& {Jardine}(2018)}]{CollierCameron2018}
{Collier Cameron}, A., \& {Jardine}, M. 2018, \mnras, 476, 2542,
  \dodoi{10.1093/mnras/sty292}

\bibitem[{Collins {et~al.}(2017)Collins, Kielkopf, \& Stassun}]{Collins2017}
Collins, K.~A., Kielkopf, J.~F., \& Stassun, K.~G. 2017, The Astronomical
  Journal, 153, 78, \dodoi{10.3847/1538-3881/153/2/78}

\bibitem[{Copperwheat {et~al.}(2013)Copperwheat, Wheatley, Southworth, Bento,
  Marsh, Dhillon, Fortney, Littlefair, \& Hickman}]{Copperwheat2013}
Copperwheat, C.~M., Wheatley, P.~J., Southworth, J., {et~al.} 2013, Monthly
  Notices of the Royal Astronomical Society, 434, 661,
  \dodoi{10.1093/mnras/stt1056}

\bibitem[{Counselman(1973)}]{Counselman1973}
Counselman, C.~C. 1973, The Astrophysical Journal, 180, 307,
  \dodoi{10.1086/151964}

\bibitem[{Cowan {et~al.}(2012)Cowan, Machalek, Croll, Shekhtman, Burrows,
  Deming, Greene, \& Hora}]{Cowan2012}
Cowan, N.~B., Machalek, P., Croll, B., {et~al.} 2012, The Astrophysical
  Journal, 747, 82, \dodoi{10.1088/0004-637X/747/1/82}

\bibitem[{Croll {et~al.}(2015)Croll, Albert, Jayawardhana, Cushing, Moutou,
  Lafreniere, Johnson, Bonomo, Deleuil, \& Fortney}]{Croll2015}
Croll, B., Albert, L., Jayawardhana, R., {et~al.} 2015, The Astrophysical
  Journal, 802, 28, \dodoi{10.1088/0004-637X/802/1/28}

\bibitem[{Crossfield {et~al.}(2012)Crossfield, Barman, Hansen, Tanaka, \&
  Kodama}]{Crossfield2012}
Crossfield, I. J.~M., Barman, T., Hansen, B. M.~S., Tanaka, I., \& Kodama, T.
  2012, The Astrophysical Journal, 760, 140,
  \dodoi{10.1088/0004-637X/760/2/140}

\bibitem[{Csizmadia {et~al.}(2019)Csizmadia, Hellard, \& Smith}]{Csizmadia2019}
Csizmadia, S., Hellard, H., \& Smith, A. M.~S. 2019, Astronomy \& Astrophysics,
  623, A45, \dodoi{10.1051/0004-6361/201834376}

\bibitem[{Deming {et~al.}(2015)Deming, Knutson, Kammer, Fulton, Ingalls, Carey,
  Burrows, Fortney, Todorov, Agol, Cowan, Desert, Fraine, Langton, Morley, \&
  Showman}]{Deming2015}
Deming, D., Knutson, H., Kammer, J., {et~al.} 2015, The Astrophysical Journal,
  805, 132, \dodoi{10.1088/0004-637X/805/2/132}

\bibitem[{{Eastman} {et~al.}(2013){Eastman}, {Gaudi}, \& {Agol}}]{Eastman2013}
{Eastman}, J., {Gaudi}, B.~S., \& {Agol}, E. 2013, \pasp, 125, 83,
  \dodoi{10.1086/669497}

\bibitem[{Essick \& Weinberg(2015)}]{Essick2015}
Essick, R., \& Weinberg, N.~N. 2015, The Astrophysical Journal, 816, 18,
  \dodoi{10.3847/0004-637X/816/1/18}

\bibitem[{F{\"o}hring {et~al.}(2013)F{\"o}hring, Dhillon, Madhusudhan, Marsh,
  Copperwheat, Littlefair, \& Wilson}]{Fohring2013}
F{\"o}hring, D., Dhillon, V.~S., Madhusudhan, N., {et~al.} 2013, Monthly
  Notices of the Royal Astronomical Society, 435, 2268,
  \dodoi{10.1093/mnras/stt1443}

\bibitem[{{Foreman-Mackey} {et~al.}(2013){Foreman-Mackey}, {Hogg}, {Lang}, \&
  {Goodman}}]{Foreman-Mackey13}
{Foreman-Mackey}, D., {Hogg}, D.~W., {Lang}, D., \& {Goodman}, J. 2013, \pasp,
  125, 306, \dodoi{10.1086/670067}

\bibitem[{Fossati {et~al.}(2010)Fossati, Haswell, Froning, Hebb, Holmes, Kolb,
  Helling, Carter, Wheatley, Cameron, Loeillet, Pollacco, Street, Stempels,
  Simpson, Udry, Joshi, West, Skillen, \& Wilson}]{Fossati2010}
Fossati, L., Haswell, C.~A., Froning, C.~S., {et~al.} 2010, The Astrophysical
  Journal, 714, L222, \dodoi{10.1088/2041-8205/714/2/L222}

\bibitem[{{Fulton} {et~al.}(2018){Fulton}, {Petigura}, {Blunt}, \&
  {Sinukoff}}]{Fulton18}
{Fulton}, B.~J., {Petigura}, E.~A., {Blunt}, S., \& {Sinukoff}, E. 2018, \pasp,
  130, 044504, \dodoi{10.1088/1538-3873/aaaaa8}

\bibitem[{{Gelman} \& {Rubin}(1992)}]{GelmanRubin}
{Gelman}, A., \& {Rubin}, D.~B. 1992, Statistical Science, 7, 457,
  \dodoi{10.1214/ss/1177011136}

\bibitem[{Geweke(1992)}]{Geweke1992}
Geweke, J. 1992, Statist. Sci., 7, 94, \dodoi{10.1214/ss/1177011446}

\bibitem[{Gim{\'e}nez \& Bastero(1995)}]{Gimenez1995}
Gim{\'e}nez, A., \& Bastero, M. 1995, Astrophysics and Space Science, 226, 99,
  \dodoi{10.1007/BF00626903}

\bibitem[{Goldreich \& Soter(1966)}]{Goldreich1966}
Goldreich, P., \& Soter, S. 1966, Icarus, 5, 375,
  \dodoi{10.1016/0019-1035(66)90051-0}

\bibitem[{Goodman \& Weare(2010)}]{Goodman10}
Goodman, J., \& Weare, J. 2010, Communications in Applied Mathematics and
  Computational Science, 5, 65, \dodoi{10.2140/camcos.2010.5.65}

\bibitem[{Hamer \& Schlaufman(2019)}]{Hamer2019}
Hamer, J.~H., \& Schlaufman, K.~C. 2019, arXiv:1908.06998 [astro-ph].
\newblock \doarXiv{1908.06998}

\bibitem[{Haswell(2018)}]{Haswell2018}
Haswell, C.~A. 2018, in Handbook of {{Exoplanets}}, ed. H.~J. Deeg \& J.~A.
  Belmonte ({Cham}: {Springer International Publishing}), 2585--2602,
  \dodoi{10.1007/978-3-319-55333-7_97}

\bibitem[{Haswell {et~al.}(2012)Haswell, Fossati, Ayres, France, Froning,
  Holmes, Kolb, Busuttil, Street, Hebb, Cameron, Enoch, Burwitz, Rodriguez,
  West, Pollacco, Wheatley, \& Carter}]{Haswell2012}
Haswell, C.~A., Fossati, L., Ayres, T., {et~al.} 2012, The Astrophysical
  Journal, 760, 79, \dodoi{10.1088/0004-637X/760/1/79}

\bibitem[{Hebb {et~al.}(2009)Hebb, {Collier-Cameron}, Loeillet, Pollacco,
  H{\'e}brard, Street, Bouchy, Stempels, Moutou, Simpson, Udry, Joshi, West,
  Skillen, Wilson, McDonald, Gibson, Aigrain, Anderson, Benn, Christian, Enoch,
  Haswell, Hellier, Horne, Irwin, Lister, Maxted, Mayor, Norton, Parley, Pont,
  Queloz, Smalley, \& Wheatley}]{Hebb2009}
Hebb, L., {Collier-Cameron}, A., Loeillet, B., {et~al.} 2009, The Astrophysical
  Journal, 693, 1920, \dodoi{10.1088/0004-637X/693/2/1920}

\bibitem[{{Holman} {et~al.}(1997){Holman}, {Touma}, \& {Tremaine}}]{Holman1997}
{Holman}, M., {Touma}, J., \& {Tremaine}, S. 1997, \nat, 386, 254,
  \dodoi{10.1038/386254a0}

\bibitem[{Hooton {et~al.}(2019)Hooton, {de Mooij}, Watson, Gibson,
  {Galindo-Guil}, Clavero, \& Merritt}]{Hooton2019}
Hooton, M.~J., {de Mooij}, E. J.~W., Watson, C.~A., {et~al.} 2019, Monthly
  Notices of the Royal Astronomical Society, 486, 2397,
  \dodoi{10.1093/mnras/stz966}

\bibitem[{Howard {et~al.}(2010)Howard, Johnson, Marcy, Fischer, Wright, Bernat,
  Henry, Peek, Isaacson, Apps, Endl, Cochran, Valenti, Anderson, \&
  Piskunov}]{Howard2010}
Howard, A.~W., Johnson, J.~A., Marcy, G.~W., {et~al.} 2010, The Astrophysical
  Journal, 721, 1467, \dodoi{10.1088/0004-637X/721/2/1467}

\bibitem[{Huang {et~al.}(2016)Huang, Wu, \& Triaud}]{Huang2016}
Huang, C., Wu, Y., \& Triaud, A. H. M.~J. 2016, The Astrophysical Journal, 825,
  98, \dodoi{10.3847/0004-637X/825/2/98}

\bibitem[{Husnoo {et~al.}(2012)Husnoo, Pont, Mazeh, Fabrycky, H{\'e}brard,
  Bouchy, \& Shporer}]{Husnoo2012}
Husnoo, N., Pont, F., Mazeh, T., {et~al.} 2012, Monthly Notices of the Royal
  Astronomical Society, 422, 3151, \dodoi{10.1111/j.1365-2966.2012.20839.x}

\bibitem[{Husnoo {et~al.}(2011)Husnoo, Pont, H{\'e}brard, Simpson, Mazeh,
  Bouchy, Moutou, Arnold, Boisse, D{\'i}az, Eggenberger, \&
  Shporer}]{Husnoo2011}
Husnoo, N., Pont, F., H{\'e}brard, G., {et~al.} 2011, Monthly Notices of the
  Royal Astronomical Society, 413, 2500,
  \dodoi{10.1111/j.1365-2966.2011.18322.x}

\bibitem[{Hut(1980)}]{Hut1980}
Hut, P. 1980, Astronomy and Astrophysics, 92, 167

\bibitem[{{Innanen} {et~al.}(1997){Innanen}, {Zheng}, {Mikkola}, \&
  {Valtonen}}]{Innanen1997}
{Innanen}, K.~A., {Zheng}, J.~Q., {Mikkola}, S., \& {Valtonen}, M.~J. 1997,
  \aj, 113, 1915, \dodoi{10.1086/118405}

\bibitem[{Jackson {et~al.}(2017)Jackson, Arras, Penev, Peacock, \&
  Marchant}]{Jackson2017}
Jackson, B., Arras, P., Penev, K., Peacock, S., \& Marchant, P. 2017, The
  Astrophysical Journal, 835, 145, \dodoi{10.3847/1538-4357/835/2/145}

\bibitem[{Jackson {et~al.}(2008)Jackson, Greenberg, \& Barnes}]{Jackson2008}
Jackson, B., Greenberg, R., \& Barnes, R. 2008, The Astrophysical Journal, 678,
  1396, \dodoi{10.1086/529187}

\bibitem[{Jackson {et~al.}(2016)Jackson, Jensen, Peacock, Arras, \&
  Penev}]{Jackson2016}
Jackson, B., Jensen, E., Peacock, S., Arras, P., \& Penev, K. 2016, Celestial
  Mechanics and Dynamical Astronomy, 126, 227,
  \dodoi{10.1007/s10569-016-9704-1}

\bibitem[{Jia \& Spruit(2017)}]{Jia2017}
Jia, S., \& Spruit, H.~C. 2017, Monthly Notices of the Royal Astronomical
  Society, 465, 149, \dodoi{10.1093/mnras/stw1693}

\bibitem[{Jones {et~al.}(2001)Jones, Oliphant, Peterson, {et~al.}}]{Scipy}
Jones, E., Oliphant, T., Peterson, P., {et~al.} 2001, {SciPy}: Open source
  scientific tools for {Python}.
\newblock \url{http://www.scipy.org/}

\bibitem[{Jord{\'a}n \& Bakos(2008)}]{Jordan2008}
Jord{\'a}n, A., \& Bakos, G.~{\'A}. 2008, The Astrophysical Journal, 685, 543,
  \dodoi{10.1086/590549}

\bibitem[{Knutson {et~al.}(2014)Knutson, Fulton, Montet, Kao, Ngo, Howard,
  Crepp, Hinkley, Bakos, Batygin, Johnson, Morton, \& Muirhead}]{Knutson2014}
Knutson, H.~A., Fulton, B.~J., Montet, B.~T., {et~al.} 2014, The Astrophysical
  Journal, 785, 126, \dodoi{10.1088/0004-637X/785/2/126}

\bibitem[{Kreidberg(2015)}]{Kreidberg2015}
Kreidberg, L. 2015, Publications of the Astronomical Society of the Pacific,
  127, 1161, \dodoi{10.1086/683602}

\bibitem[{Kreidberg {et~al.}(2015)Kreidberg, Line, Bean, Stevenson, D{\'e}sert,
  Madhusudhan, Fortney, Barstow, Henry, Williamson, \&
  Showman}]{Kreidberg2015a}
Kreidberg, L., Line, M.~R., Bean, J.~L., {et~al.} 2015, The Astrophysical
  Journal, 814, 66, \dodoi{10.1088/0004-637X/814/1/66}

\bibitem[{Lai {et~al.}(2010)Lai, Helling, \& van~den Heuvel}]{Lai2010}
Lai, D., Helling, C., \& van~den Heuvel, E. P.~J. 2010, The Astrophysical
  Journal, 721, 923, \dodoi{10.1088/0004-637X/721/2/923}

\bibitem[{Levrard {et~al.}(2009)Levrard, Winisdoerffer, \&
  Chabrier}]{Levrard2009}
Levrard, B., Winisdoerffer, C., \& Chabrier, G. 2009, The Astrophysical
  Journal, 692, L9, \dodoi{10.1088/0004-637X/692/1/L9}

\bibitem[{{Luger} {et~al.}(2016){Luger}, {Agol}, {Kruse}, {Barnes}, {Becker},
  {Foreman-Mackey}, \& {Deming}}]{Luger2016}
{Luger}, R., {Agol}, E., {Kruse}, E., {et~al.} 2016, \aj, 152, 100,
  \dodoi{10.3847/0004-6256/152/4/100}

\bibitem[{Maciejewski {et~al.}(2011)Maciejewski, Errmann, Raetz, Seeliger,
  Spaleniak, \& Neuh{\"a}user}]{Maciejewski2011}
Maciejewski, G., Errmann, R., Raetz, S., {et~al.} 2011, Astronomy \&
  Astrophysics, 528, A65, \dodoi{10.1051/0004-6361/201016268}

\bibitem[{Maciejewski {et~al.}(2013)Maciejewski, Dimitrov, Seeliger, Raetz,
  Bukowiecki, Kitze, Errmann, Nowak, Niedzielski, Popov, Marka,
  Go{\'z}dziewski, Neuh{\"a}user, Ohlert, Hinse, Lee, Lee, Yoon, Berndt,
  Gilbert, Ginski, Hohle, Mugrauer, R{\"o}ll, Schmidt, Tetzlaff, Mancini,
  Southworth, Dall'Ora, Ciceri, Zambelli, Corfini, Takahashi, Tachihara,
  Benk{\H o}, S{\'a}rneczky, Szabo, Varga, Va{\v n}ko, Joshi, \&
  Chen}]{Maciejewski2013}
Maciejewski, G., Dimitrov, D., Seeliger, M., {et~al.} 2013, Astronomy \&
  Astrophysics, 551, A108, \dodoi{10.1051/0004-6361/201220739}

\bibitem[{Maciejewski {et~al.}(2016)Maciejewski, Dimitrov, Fern{\'a}ndez, Sota,
  Nowak, Ohlert, Nikolov, Bukowiecki, Hinse, Pall{\'e}, Tingley, Kjurkchieva,
  Lee, \& Lee}]{Maciejewski2016}
Maciejewski, G., Dimitrov, D., Fern{\'a}ndez, M., {et~al.} 2016, Astronomy \&
  Astrophysics, 588, L6, \dodoi{10.1051/0004-6361/201628312}

\bibitem[{Maciejewski {et~al.}(2018)Maciejewski, Fern{\'a}ndez, Aceituno,
  {Mart{\`i}n-Ruiz}, Ohlert, Dimitrov, Szyszka, {von Essen}, Mugrauer,
  Bischoff, Michel, Mallonn, Stangret, \& Mo{\'z}dzierski}]{Maciejewski2018}
Maciejewski, G., Fern{\'a}ndez, M., Aceituno, F., {et~al.} 2018, Acta
  Astronomica, 68, 371, \dodoi{10.32023/0001-5237/68.4.4}

\bibitem[{{Mandel} \& {Agol}(2002)}]{Mandel02}
{Mandel}, K., \& {Agol}, E. 2002, \apjl, 580, L171, \dodoi{10.1086/345520}

\bibitem[{{Mazeh} {et~al.}(1997){Mazeh}, {Krymolowski}, \&
  {Rosenfeld}}]{Mazeh1997}
{Mazeh}, T., {Krymolowski}, Y., \& {Rosenfeld}, G. 1997, \apjl, 477, L103,
  \dodoi{10.1086/310536}

\bibitem[{Meibom \& Mathieu(2005)}]{Meibom2005}
Meibom, S., \& Mathieu, R.~D. 2005, The Astrophysical Journal, 620, 970,
  \dodoi{10.1086/427082}

\bibitem[{Millholland \& Laughlin(2018)}]{Millholland2018}
Millholland, S., \& Laughlin, G. 2018, The Astrophysical Journal, 869, L15,
  \dodoi{10.3847/2041-8213/aaedb1}

\bibitem[{{Miralda-Escud{\'e}}(2002)}]{Miralda-Escude2002}
{Miralda-Escud{\'e}}, J. 2002, The Astrophysical Journal, 564, 1019,
  \dodoi{10.1086/324279}

\bibitem[{Nichols {et~al.}(2015)Nichols, Wynn, Goad, Alexander, Casewell,
  Cowley, Burleigh, Clarke, \& Bisikalo}]{Nichols2015}
Nichols, J.~D., Wynn, G.~A., Goad, M., {et~al.} 2015, The Astrophysical
  Journal, 803, 9, \dodoi{10.1088/0004-637X/803/1/9}

\bibitem[{Ogilvie(2014)}]{Ogilvie2014}
Ogilvie, G.~I. 2014, Annual Review of Astronomy and Astrophysics, 52, 171,
  \dodoi{10.1146/annurev-astro-081913-035941}

\bibitem[{{\"O}zt{\"u}rk \& Erdem(2019)}]{Ozturk2019}
{\"O}zt{\"u}rk, O., \& Erdem, A. 2019, Monthly Notices of the Royal
  Astronomical Society, 486, 2290, \dodoi{10.1093/mnras/stz747}

\bibitem[{P{\'a}l \& Kocsis(2008)}]{Pal2008}
P{\'a}l, A., \& Kocsis, B. 2008, Monthly Notices of the Royal Astronomical
  Society, 389, 191, \dodoi{10.1111/j.1365-2966.2008.13512.x}

\bibitem[{Patra {et~al.}(2017)Patra, Winn, Holman, Yu, Deming, \&
  Dai}]{Patra2017}
Patra, K.~C., Winn, J.~N., Holman, M.~J., {et~al.} 2017, The Astronomical
  Journal, 154, 4, \dodoi{10.3847/1538-3881/aa6d75}

\bibitem[{Penev {et~al.}(2018)Penev, Bouma, Winn, \& Hartman}]{Penev2018}
Penev, K., Bouma, L.~G., Winn, J.~N., \& Hartman, J.~D. 2018, The Astronomical
  Journal, 155, 165, \dodoi{10.3847/1538-3881/aaaf71}

\bibitem[{Penev {et~al.}(2012)Penev, Jackson, Spada, \& Thom}]{Penev2012}
Penev, K., Jackson, B., Spada, F., \& Thom, N. 2012, The Astrophysical Journal,
  751, 96, \dodoi{10.1088/0004-637X/751/2/96}

\bibitem[{{Petrucci} {et~al.}(2019){Petrucci}, {Jofr{\'e}}, {G{\'o}mez Maqueo
  Chew}, {Hinse}, {Ma{\v{s}}ek}, {Tan}, \& {G{\'o}mez}}]{Petrucci2019}
{Petrucci}, R., {Jofr{\'e}}, E., {G{\'o}mez Maqueo Chew}, Y., {et~al.} 2019,
  \mnras, 2628, \dodoi{10.1093/mnras/stz3034}

\bibitem[{Ragozzine \& Wolf(2009)}]{Ragozzine2009}
Ragozzine, D., \& Wolf, A.~S. 2009, The Astrophysical Journal, 698, 1778,
  \dodoi{10.1088/0004-637X/698/2/1778}

\bibitem[{Rappaport {et~al.}(2013)Rappaport, {Sanchis-Ojeda}, Rogers, Levine,
  \& Winn}]{Rappaport2013}
Rappaport, S., {Sanchis-Ojeda}, R., Rogers, L.~A., Levine, A., \& Winn, J.~N.
  2013, The Astrophysical Journal, 773, L15,
  \dodoi{10.1088/2041-8205/773/1/L15}

\bibitem[{Rasio {et~al.}(1996)Rasio, Tout, Lubow, \& Livio}]{Rasio1996}
Rasio, F.~A., Tout, C.~A., Lubow, S.~H., \& Livio, M. 1996, The Astrophysical
  Journal, 470, 1187, \dodoi{10.1086/177941}

\bibitem[{{Ricker} {et~al.}(2014){Ricker}, {Winn}, {Vanderspek}, {Latham},
  {Bakos}, {Bean}, {Berta-Thompson}, {Brown}, {Buchhave}, {Butler}, {Butler},
  {Chaplin}, {Charbonneau}, {Christensen-Dalsgaard}, {Clampin}, {Deming},
  {Doty}, {De Lee}, {Dressing}, {Dunham}, {Endl}, {Fressin}, {Ge}, {Henning},
  {Holman}, {Howard}, {Ida}, {Jenkins}, {Jernigan}, {Johnson}, {Kaltenegger},
  {Kawai}, {Kjeldsen}, {Laughlin}, {Levine}, {Lin}, {Lissauer}, {MacQueen},
  {Marcy}, {McCullough}, {Morton}, {Narita}, {Paegert}, {Palle}, {Pepe},
  {Pepper}, {Quirrenbach}, {Rinehart}, {Sasselov}, {Sato}, {Seager},
  {Sozzetti}, {Stassun}, {Sullivan}, {Szentgyorgyi}, {Torres}, {Udry}, \&
  {Villasenor}}]{Ricker14}
{Ricker}, G.~R., {Winn}, J.~N., {Vanderspek}, R., {et~al.} 2014, in \procspie,
  Vol. 9143, Space Telescopes and Instrumentation 2014: Optical, Infrared, and
  Millimeter Wave, 914320, \dodoi{10.1117/12.2063489}

\bibitem[{Sada {et~al.}(2012)Sada, Deming, Jennings, k.~Jackson, Hamilton,
  Fraine, Peterson, Haase, Bays, Lunsford, \& O'Gorman}]{Sada2012}
Sada, P.~V., Deming, D., Jennings, D.~E., {et~al.} 2012, Publications of the
  Astronomical Society of the Pacific, 124, 212, \dodoi{10.1086/665043}

\bibitem[{Sasselov(2003)}]{Sasselov2003}
Sasselov, D.~D. 2003, The Astrophysical Journal, 596, 1327,
  \dodoi{10.1086/378145}

\bibitem[{Schwarz(1978)}]{Schwarz1978}
Schwarz, G. 1978, The Annals of Statistics, 6, 461,
  \dodoi{10.1214/aos/1176344136}

\bibitem[{Stefansson {et~al.}(2017)Stefansson, Mahadevan, Hebb, Wisniewski,
  Huehnerhoff, Morris, Halverson, Zhao, Wright, O'rourke, Knutson, Hawley,
  Kanodia, Li, Hagen, Liu, Beatty, Bender, Robertson, Dembicky, Gray,
  Ketzeback, McMillan, \& Rudyk}]{Stefansson2017}
Stefansson, G., Mahadevan, S., Hebb, L., {et~al.} 2017, The Astrophysical
  Journal, 848, 9, \dodoi{10.3847/1538-4357/aa88aa}

\bibitem[{Steffen {et~al.}(2012)Steffen, Ragozzine, Fabrycky, Carter, Ford,
  Holman, Rowe, Welsh, Borucki, Boss, Ciardi, \& Quinn}]{Steffen2012}
Steffen, J.~H., Ragozzine, D., Fabrycky, D.~C., {et~al.} 2012, Proceedings of
  the National Academy of Sciences, 109, 7982, \dodoi{10.1073/pnas.1120970109}

\bibitem[{Stevenson {et~al.}(2014)Stevenson, Bean, Seifahrt, D{\'e}sert,
  Madhusudhan, Bergmann, Kreidberg, \& Homeier}]{Stevenson2014}
Stevenson, K.~B., Bean, J.~L., Seifahrt, A., {et~al.} 2014, The Astronomical
  Journal, 147, 161, \dodoi{10.1088/0004-6256/147/6/161}

\bibitem[{Tinyanont {et~al.}(2019)Tinyanont, {Millar-Blanchaer}, Nilsson,
  Mawet, Knutson, Kataria, Vasisht, Henderson, Matthews, Serabyn, Milburn,
  Hale, Smith, Vissapragada, Santos~Jr, Kekas, \& Escuti}]{Tinyanont2019}
Tinyanont, S., {Millar-Blanchaer}, M.~A., Nilsson, R., {et~al.} 2019,
  Publications of the Astronomical Society of the Pacific, 131, 025001,
  \dodoi{10.1088/1538-3873/aaef0f}

\bibitem[{Valsecchi {et~al.}(2015)Valsecchi, Rappaport, Rasio, Marchant, \&
  Rogers}]{Valsecchi2015}
Valsecchi, F., Rappaport, S., Rasio, F.~A., Marchant, P., \& Rogers, L.~A.
  2015, The Astrophysical Journal, 813, 101,
  \dodoi{10.1088/0004-637X/813/2/101}

\bibitem[{{Valsecchi} \& {Rasio}(2014)}]{Valsecchi2014b}
{Valsecchi}, F., \& {Rasio}, F.~A. 2014, \apjl, 787, L9,
  \dodoi{10.1088/2041-8205/787/1/L9}

\bibitem[{{Van Der Walt} {et~al.}(2011){Van Der Walt}, {Colbert}, \&
  {Varoquaux}}]{Numpy}
{Van Der Walt}, S., {Colbert}, S.~C., \& {Varoquaux}, G. 2011, ArXiv e-prints.
\newblock \doarXiv{1102.1523}

\bibitem[{Vissapragada {et~al.}(2019)Vissapragada, {Jontof-Hutter}, Shporer,
  Knutson, Liu, Thorngren, Lee, Chachan, Mawet, {Millar-Blanchaer}, Nilsson,
  Tinyanont, Vasisht, \& Wright}]{Vissapragada2019}
Vissapragada, S., {Jontof-Hutter}, D., Shporer, A., {et~al.} 2019,
  arXiv:1907.04445 [astro-ph].
\newblock \doarXiv{1907.04445}

\bibitem[{{Vogt} {et~al.}(1994){Vogt}, {Allen}, {Bigelow}, {Bresee}, {Brown},
  {Cantrall}, {Conrad}, {Couture}, {Delaney}, {Epps}, {Hilyard}, {Hilyard},
  {Horn}, {Jern}, {Kanto}, {Keane}, {Kibrick}, {Lewis}, {Osborne},
  {Pardeilhan}, {Pfister}, {Ricketts}, {Robinson}, {Stover}, {Tucker}, {Ward},
  \& {Wei}}]{Vogt94}
{Vogt}, S.~S., {Allen}, S.~L., {Bigelow}, B.~C., {et~al.} 1994, 2198, 362

\bibitem[{{von Essen} {et~al.}(2019){von Essen}, Stefansson, Mallonn, Pursimo,
  Djupvik, Mahadevan, Kjeldsen, Freudenthal, \& Dreizler}]{vonEssen2019}
{von Essen}, C., Stefansson, G., Mallonn, M., {et~al.} 2019, arXiv:1904.05362
  [astro-ph].
\newblock \doarXiv{1904.05362}

\bibitem[{Weinberg {et~al.}(2017)Weinberg, Sun, Arras, \&
  Essick}]{Weinberg2017}
Weinberg, N.~N., Sun, M., Arras, P., \& Essick, R. 2017, The Astrophysical
  Journal, 849, L11, \dodoi{10.3847/2041-8213/aa9113}

\bibitem[{Wilson {et~al.}(2003)Wilson, Eikenberry, Henderson, Hayward, Carson,
  Pirger, Barry, Brandl, Houck, Fitzgerald, \& Stolberg}]{Wilson2003a}
Wilson, J.~C., Eikenberry, S.~S., Henderson, C.~P., {et~al.} 2003, in
  Instrument {{Design}} and {{Performance}} for {{Optical}}/{{Infrared
  Ground}}-Based {{Telescopes}}, Vol. 4841 ({International Society for Optics
  and Photonics}), 451--458, \dodoi{10.1117/12.460336}

\bibitem[{Winn(2010)}]{Winn2010}
Winn, J.~N. 2010, in Exoplanets, Edited by {{S}}. {{Seager}}. {{Published}} by
  {{University}} of {{Arizona Press}}, {{Tucson}}, {{AZ}}, 2010, 526 Pp.
  {{ISBN}} 978-0-8165-2945-2., p.55-77, ed. S.~Seager, 55--77.
\newblock \doarXiv{1001.2010}

\end{thebibliography}
\bibliographystyle{aasjournal}

\end{document}